\documentclass[%
reprint,
superscriptaddress,
preprintnumbers,
 amsmath,amssymb,
 aps,
]{revtex4-1}

\pdfoutput=1

\usepackage{graphicx}
\usepackage{dcolumn}
\usepackage{bm}
\usepackage{slashed}
\usepackage[colorlinks=true, pdfstartview=FitV, linkcolor=red, citecolor=blue, urlcolor=blue]{hyperref}
\usepackage{slashed}

\newcommand{\bs}{\boldsymbol}
\newcommand{\be}{\begin{equation}}      
\newcommand{\ee}{\end{equation}}      
\newcommand{\bea}{\begin{eqnarray}}      
\newcommand{\eea}{\end{eqnarray}}

\newcommand{\cH}{{\cal H}}
\newcommand{\vphi}{\varphi}
\newcommand{\ve}{\varepsilon}
\newcommand{\cA}{\mathcal{A}}
\newcommand{\dif}{\mathrm{d}}

\usepackage[normalem]{ulem}
\renewcommand\sout{\bgroup \color{red} \ULdepth=-.5ex \ULset}

\begin{document}

\preprint{RBRC 1231}

\title{Topological properties of the chiral magnetic effect in multi-Weyl semimetals}

\author{Tomoya Hayata}
\email[]{hayata@phys.chuo-u.ac.jp}
\affiliation{Department of Physics, Chuo University, 1-13-27 Kasuga, Bunkyo, Tokyo 112-8551, Japan}

\author{Yuta Kikuchi}
\email[]{yuta.kikuchi@stonybrook.edu}
\affiliation{Department of Physics and Astronomy, Stony Brook University, Stony Brook, New York 11794-3800, USA}
\affiliation{Department of Physics, Kyoto University, Kyoto 606-8502, Japan}

\author{Yuya Tanizaki}
\email[]{yuya.tanizaki@riken.jp}
\affiliation{RIKEN BNL Research Center, Brookhaven National Laboratory, Upton, NY 11973-5000 USA}

\date{\today}

\begin{abstract}
 We compute the chiral magnetic effect (CME) in multi-Weyl semimetals (multi-WSMs) based on the chiral kinetic theory. 
Multi-WSMs are WSMs with multiple monopole charges that have nonlinear and anisotropic dispersion relations near Weyl points, and we need to extend conventional computation of CME in WSMs with linear dispersion relations.  
Topological properties of CME in multi-WSMs are investigated in details for not only static magnetic fields but also time-dependent (dynamic) ones. 
We propose an experimental setup to measure the multiple monopole charge via the topological nature hidden in the dynamic CME.
\end{abstract}

\maketitle



\section{\label{sec:1}
Introduction}

Weyl semimetals (WSMs), a family of the topological materials, possess band touching points characterized by the nontrivial topology (monopole charge) in momentum space~\cite{Murakami,PhysRevB.83.205101,PhysRevLett.107.127205,Xu613,Lu622,PhysRevX.5.031013}. Excitations of electrons near those points share a lot of properties in common with the (3+1)-dimensional relativistic Weyl fermions in particle physics. Henceforth, quantum field theory of relativistic Weyl fermions is directly applicable to WSMs and, in particular, the quantum anomaly plays an important role in the description of transport phenomena in WSMs. 
%
The chiral magnetic effect (CME)~\cite{Nielsen:1983rb,Fukushima:2008xe} is one of the outstanding transport phenomena realized in WSMs as well as Dirac semimetals. Experimental measurements have confirmed a key signal of CME in Weyl/Dirac semimetals, that is, the negative and anisotropic magnetoresistance~\cite{PhysRevB.88.104412}, over the past few years~\cite{KharzeevDirac,xiong2015evidence,li2015giant,huang2015observation,wang2015helicity,zhang2015observation,shekhar2015large}.

While similarity to relativistic systems has driven the theoretical development of WSMs, condensed matter systems realizing WSMs generally do not have the Lorentz symmetry nor continuous rotational symmetry. 
Such a fact leads to theoretical prediction of new types of Weyl fermions unique to condensed matter systems 
and has drawn further attention in recent years. 
One of the possibilities is the WSMs with multiple monopole charge and
they are called multi-WSMs \cite{PhysRevLett.107.186806,fang2012multi,PhysRevB.93.155125,park2017semiclassical}. The multiple monopole charge requires nonlinear and anisotropic dispersion relations near band-touching points. 
It was pointed out that multi-WSMs are protected by point-group symmetries~\cite{fang2012multi}. 
For instance, $C_4$ and $C_6$ discrete rotational symmetries yield quadratic and cubic band touchings with the double and triple monopole charges, respectively. Although the existence of this type of topologically protected state is promising, systematic studies on transport phenomena in multi-WSMs are not yet completely done.

We will make a clear distinction between two different kinds of CME, static and dynamic CMEs, throughout this paper. Static CME refers CME under time-independent magnetic fields, and dynamic CME refers CME under time-dependent magnetic fields.
For static CME, there had been debates on the existence of chiral magnetic current in equilibirum, which is dissipationless, and there seems to be no energy source for the induction of current~\cite{PhysRevLett.111.027201,PhysRevB.88.125105,Basar:2013iaa,PhysRevB.89.075124,PhysRevD.92.085011,PhysRevB.91.115203}. 
It is now understood that the static CME necessarily requires ``chiral chemical potential'', which is actually induced only by driving the system into nonequilibrium states.
On the other hand, the dynamic CME is the AC current induced by time-dependent magnetic fields~\cite{RevModPhys.82.1959,PhysRevD.87.085016,PhysRevB.94.115160,zhong2016gyrotropic,kharzeev2016anatomy}. Since the dynamical magnetic fields drives the systems out of equilibrium, the dynamic CME can occur even in the absence of chiral chemical potential. 
This is expected to be observed in optical measurement~\cite{PhysRevLett.118.107401,zhong2016gyrotropic}.

In this paper, we investigate the static and dynamic CMEs in multi-WSMs with focusing on their topological properties, based on the chiral kinetic theory~\cite{Stephanov:2012ki,Son:2012wh,Chen:2012ca}, i.e., the kinetic theory incorporating the effect of the Berry curvature~\cite{PhysRevB.59.14915,RevModPhys.82.1959}.  
In conventional WSMs with linear dispersion relations and unit monopole charges, CME under static external magnetic fields $\bs{B}$ (static CME) is given by $\bs{J}_\mathrm{CME}=(e^2/2\pi^2)\mu_5\bs{B}$, where $e$ and $\mu_5$ are the unit charge and chiral chemical potential~\cite{Fukushima:2008xe} \footnote{In this paper, we define the chiral chemical potential $\mu_5$ in such a way that $\mu_5$ vanishes in the equilibrium state even if there is a energy separation $b_0$ between left and right handed Weyl cones.}. This expression is robust against thermal excitations as it does not depend on temperature.
In order to extend it to the case with the multiple monopole charge, we need to consider nonlinear and anisotropic band touching points that cause a lot of technical complications, especially at finite temperature. We introduce the general technique to compute topological quantities of multi-WSMs in an efficient way, and explicitly show that the above expression is just multiplied by the monopole charge $K$ \cite{Son:2012wh}:
\begin{align}
\label{eq:intro_CME}
\bs{J}_\mathrm{CME}=\frac{e^2}{2\pi^2}K\mu_5\bs{B}.
\end{align}
The obtained result is robust against the deformation of dispersion relation and geometric deformations such as strain-induced effect or temperature gradient as well as thermal excitations (Sec.~\ref{sec:3a}).

Furthermore, we apply our computational method to the dynamic CME, where the AC current is induced in the direction of time-dependent external magnetic fields. In contrast to the static CME, the dynamic CME is \textit{not} topologically protected in anisotropic systems. However, we show that the topological nature is hidden in the dynamic CME.
More specifically, we show that the trace of the chiral magnetic conductivity exhibits the topological nature even for anisotropic systems:
\begin{align}
\label{eq:intro_dCME}
 \sum_{i=1,2,3}\sigma_\mathrm{dCME}^{ii}
 =\frac{e^2}{2\pi^2}b_0 K.
\end{align}
The chiral magnetic conductivity for dynamic CME is defined by $J^i_\mathrm{dCME}=\sigma_\mathrm{dCME}^{ij}B^j$ and $b_0$ is an energy separation between a pair of left- and right-handed Weyl nodes (see Sec.~\ref{sec:3b} for details).
This quantity is again robust against the aforementioned perturbations as well as the chiral magnetic conductivity of static CME~\eqref{eq:intro_CME}. The crucial difference between Eqs.~\eqref{eq:intro_CME} and~\eqref{eq:intro_dCME} is that the latter does not depend on chiral chemical potential but depends on $b_0$. An energy separation $b_0$ is an IR property of Weyl cones, while chiral chemical potential strongly depends on UV properties of the entire band structures.\footnote{The chiral chemical potential $\mu_5$ itself does not necessarily depend on UV properties of band structure. However, for the practical realization of nonzero $\mu_5$, we need some chirality source such as the chiral anomaly via external parallel electromagnetic fields $\bs{E}\cdot\bs{B}$. Then, the competition between the chirality pumping and chirality relaxation results in nonvanishing $\mu_5\propto \tau_5\bs{E}\cdot\bs{B}$ with the chirality relaxation time $\tau_5$. Therefore, $\mu_5$ indeed depends on UV properties because $\tau_5$ is determined by chirality relaxation mechanisms such as chirality flipping scattering and impurity scattering. It is very hard to theoretically predict $\tau_5$, which highly depends on microscopic nature of individual materials.}
By utilizing this advantage of dynamic CME~\eqref{eq:intro_dCME}, we may detect the monopole charge without suffering from the UV structures, which cause some subtleties on the experimental measurement of static CME. Based on these arguments,
we also address the experimental setup to detect the multiple monopole charge from measurements of the dynamic CME.
Our proposal can be tested in multi-WSMs with broken spatial-inversion and reflection symmetries such as SrSi$_2$.



\section{\label{sec:2}
Monopole charge in Multi-WSM}


In this section, we apply a useful technique to the computation of monopole charge as a demonstration. Although the result of this section is not new, the computational method enables us to explicitly show the topological properties of chiral magnetic effect in Sec.~\ref{sec:3}.

\begin{figure}
 \includegraphics[width=7cm]{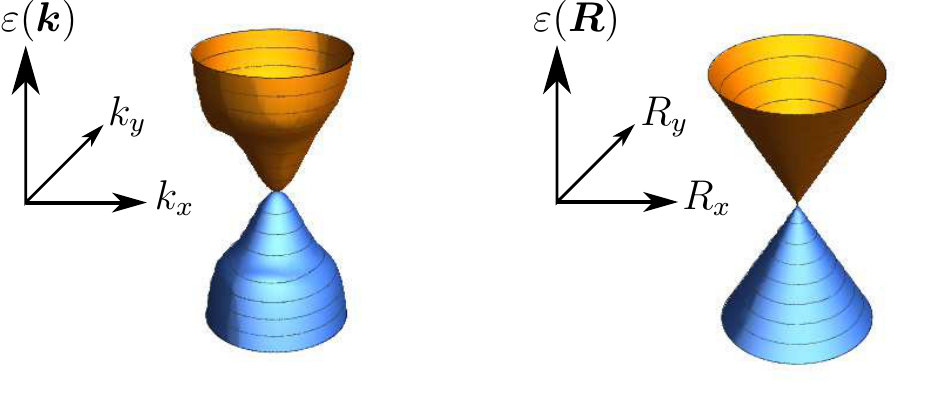}
 \caption{(Color online) The left cone is a nonlinear and anisotropic Weyl cone in $\bs{k}$-space (Brillouin zone) with $k_z=0$, described by the model Hamiltonian~\eqref{eq:mWH}. The right cone is a linear and isotropic Weyl cone in $\bs{R}$-space with $R_z=0$.}
 \label{fig:cones}
\end{figure}

We first introduce the general effective Hamiltonian describing the band structure near the two-band touching point,
\begin{align}
 \label{eq:weyl_ham}
 H=\sum_{i=x,y,z}\sigma_i R_i(\bs{k}),
\end{align}
where $\sigma_i$ are the Pauli matrices. $R_i(\bs{k})$ are some real functions with a zero only at $\bs{k}=0$. 
Here, momentum are measured from the band-touching point.
The Einstein convention is understood for repeated indices below.
The Hamiltonian describes the linear left-handed Weyl cone in $\bs{R}$ space [$\bs{R}=(R_x,R_y,R_z)$] although its shape in $\bs{k}$ space [$\bs{k}=(k_x,k_y,k_z)$] is not specified (See Fig.~\ref{fig:cones}). The only assumption is that the Jacobian $\mathrm{det}(\partial R_l/\partial k_i)$ does not vanish everywhere.
The eigenvalues of $H$ are $\ve_{\pm}(\bs{k})=\pm |\bs{R}(\bs{k})|$, and their corresponding normalized eigenvectors are 
\be
u_{\pm}(\bs{k})={1\over \sqrt{2 |\bs{R}| (|\bs{R}|\pm R_z)}}\left(\begin{array}{c}\pm |\bs{R}|+R_z\\R_x+ i R_y \end{array}\right). 
\ee
The Berry connection is defined by $ \cA_\pm \equiv -i u_\pm^\dagger \mathrm{d}u_\pm$, and the Berry curvature is
\begin{align}
 \label{eq:Berry_2form}
 \Omega_\pm \equiv \mathrm{d}\cA_\pm=\pm{\epsilon^{lmn}\over 4}\hat{R}_l(\bs{k})\dif \hat{R}_m(\bs{k})\wedge \dif \hat{R}_n(\bs{k}),
\end{align}
where $\hat{\bs{R}}=\bs{R}/|\bs{R}|$, and $\epsilon^{lmn}$ are completely antisymmetric tensors (See Appendix~\ref{app:0} for differential forms).
These quantities are defined in $\bs{k}$ space. 
The monopole charge is defined by the Chern number, 
\begin{align}
 K&\equiv\frac{1}{2\pi}\int_{S^2} \Omega_+, 
\label{eq:monopole}
\end{align}
where the integration is done over any two-sphere surrounding the origin $\bs{k}=0$. 
This yields the winding number characterized by $\pi_2(S^2)$ associated with a map from $\bs{k}$ space ($\mathbb{R}^3\setminus\{{0}\}$) to $\bs{R}$ space ($\mathbb{R}^3\setminus\{{0}\}$), and takes only an integer.

Hence, the multi-WSMs are described by the Hamiltonian \eqref{eq:weyl_ham} with an appropriate choice of $R_i(\bs{k})$. 
As an example, we consider a model Hamiltonian with anisotropic and polynomial dispersion relations~\cite{fang2012multi}:
\be
\cH_J=(k_-)^J\sigma_+ + (k_+)^J\sigma_- + k_z \sigma_z ,
\label{eq:mWH}
\ee
where $k_\pm=k_x\pm i k_y$, $\sigma_\pm=(\sigma_x\pm i \sigma_y)/2$, and $J$ is an integer.
To compute the monopole charge, we introduce new coordinates ($r,\theta,\vphi$) as
\be
k_{\pm} = (r\sin\theta)^{1/J}\mathrm{e}^{\pm i \vphi} ,
\ \ \ 
k_z = r\cos\theta .
\label{eq:transformation}
\ee
By using those coordinates, the Hamiltonian~\eqref{eq:mWH} reads
\be
\cH_J= r
\begin{pmatrix}
\cos\theta & \sin\theta e^{-i J\vphi} \\
\sin\theta e^{i J\vphi} & -\cos\theta
\end{pmatrix} .
\label{eq:mWHr}
\ee
The energy eigenvalues are
$\epsilon_\pm=\pm r$,
which depend only on $r$. The corresponding eigenstates are
\be
\label{eq:eigenfn_J}
u_+=
\begin{pmatrix}
\cos(\theta/2) \\
\sin(\theta/2) \mathrm{e}^{i J\vphi}
\end{pmatrix} ,
\ \ \ 
u_-=
\begin{pmatrix}
-\sin(\theta/2) \\
\cos(\theta/2) \mathrm{e}^{i J\vphi}
\end{pmatrix} .
\ee
We obtain the Berry curvature in the new coordinates as
\begin{align}
 \Omega_\pm 
 =-i\dif u_\pm^\dagger \wedge\dif u_\pm
 =\pm{1\over 2}J\sin\theta\dif\theta\wedge\dif\vphi.
\end{align}
%
The monopole charge~\eqref{eq:monopole} reads
\begin{align}
 \label{eq:mon_charge_mWSM}
 K
 =\frac{1}{2\pi}\int_{0}^{\pi} \frac{1}{2}J\sin\theta\dif\theta\int_{0}^{2\pi}\dif\vphi
 =J.
\end{align}
We see that the monopole has a multiple charge, which is determined by the power $J$ of polynomial dispersions.

\section{\label{app:1} 
Static and dynamic CME from chiral kinetic theory}

In this section, we briefly review several aspects of dynamic CME in comparison with static CME from the viewpoint of chiral kinetic theory \cite{RevModPhys.82.1959,PhysRevD.87.085016,PhysRevB.94.115160,zhong2016gyrotropic,kharzeev2016anatomy}.

\subsection{Response to static magnetic field}

We first  consider the response to static (and homogeneous) magnetic fields $\bs{B}=\mathrm{const.}$, and we set $\bs{E}=0$. 
The current under the external magnetic field takes the following form~\cite{chen2014lorentz},
\begin{align}
 \label{eq:current}
& \bs{J}=
 \nonumber\\
 &\int_{\bs{k}}\left[e\frac{\partial}{\partial\bs{k}}\left(\ve-\bs{m}\cdot\bs{B}\right)+e^2\bs{B}\left(\frac{\partial}{\partial\bs{k}}\left(\ve-\bs{m}\cdot\bs{B}\right)\cdot\bs{\Omega}\right)\right]n
 \nonumber\\
 &+\bs{\nabla}_r\times\int_{\bs{k}}\bs{m}n.
\end{align}
Here $n$ is the one-body distribution function, 
$\bs{\Omega}=(\Omega^x,\Omega^y,\Omega^z)$ are given by $\Omega^i={1\over 2}\epsilon^{ijk}\Omega^{jk}$ with $\Omega={1\over 2}\Omega^{ij}\dif k_i\wedge\dif k_j$,
and $\bs{m}$ is the magnetic moment, which is decomposed into spin and orbital parts, $\bs{m}=-(2g_s/2m_e)\bs{S}+\bs{m}^\mathrm{orb}$. 
We drop the band indices for notational simplicity in this section.
The last term is the magnetization current, which is absent in static systems as we shall see in a moment.
Under the static magnetic field, the equilibrium solution must solve the Boltzmann equation, and is given by 
\begin{align}
 n
 &=f(\ve-\bs{m}\cdot\bs{B})
 \nonumber\\
 &= f(\ve) -(\bs{m}\cdot\bs{B})\frac{\partial f(\ve)}{\partial\ve} +O(|\bs{B}|^2), 
\end{align}
where $f$ is the Fermi-Dirac distribution function. 
Since this solution is uniform, the magnetization current vanishes in the static case. 
The remaining current is calculated to be
\begin{align}
 \bs{J}= e^2\bs{B}\int_{\bs{k}}(\bs{v}\cdot\bs{\Omega})f(\ve) \equiv \bs{J}_{\mathrm{CME}},
 \label{eq:J_CME_static}
\end{align}
where $\bs{v}=\frac{\partial\ve}{\partial\bs{k}}$.
We obtain the well-known expression of static CME ~\cite{Stephanov:2012ki,Son:2012wh,Chen:2012ca}.

\subsection{Response to dynamic magnetic field}\label{sec:response_dynamic}

For dynamic magnetic field $\bs{B}\propto e^{-i\omega t+i\bs{q}\cdot\bs{r}}$, we consider the uniform limit $|\bs{q}|\to0$ by assuming that the sample size is negligibly small compared with $1/|\bs{q}|$. 
In this case, the electric and magnetic fields are tied via the Faraday law, $\frac{\partial}{\partial t}\bs{B}+\frac{\partial}{\partial\bs{r}}\times\bs{E}=0$.
In order to obtain the distribution function $n$, we solve the Boltzmann equation,
\begin{align}
 \label{eq:Boltzmann}
& \frac{\partial n}{\partial t} + \dot{\bs{r}}\cdot\frac{\partial n}{\partial\bs{r}}
+ \dot{\bs{k}}\cdot\frac{\partial n}{\partial\bs{k}}
=-\frac{n-f(\tilde{\ve})}{\tau},
\\
&(1+e\bs{B\cdot\Omega})\dot{\bs{r}}=\tilde{\bs{v}}+e\widetilde{\bs{E}}\times\bs{\Omega}+(\tilde{\bs{v}}\cdot\bs{\Omega})e\bs{B} ,
\\
&(1+e\bs{B\cdot\Omega})\dot{\bs{k}}=e\widetilde{\bs{E}}+\tilde{\bs{v}}\times e\bs{B}+(e\widetilde{\bs{E}}\cdot e\bs{B})\bs{\Omega} ,
\end{align}
where we defined $\tilde{\bs{v}}\equiv\frac{\partial\tilde{\ve}_{\bs{k}}}{\partial\bs{k}}$ and $\widetilde{\bs{E}}\equiv\bs{E}-\frac{\partial\tilde{\ve}_{\bs{k}}}{\partial\bs{r}}$. 
We introduced $\tilde{\ve}_{\bs{k}}=\ve_{\bs{k}}-\bs{m}\cdot\bs{B}$ with $\bs{m}$ being the magnetic moment  for notational simplicity.
We used the relaxation-time approximation in the right hand side of Eq.~\eqref{eq:Boltzmann}.
Substituting a decomposition of the distribution function
\begin{align}
 n=f(\ve-\bs{m}\cdot\bs{B})+\delta f,
\end{align}
into Eq.~\eqref{eq:Boltzmann}, and keeping the terms up to linear order in $\bs{E}$ and $\bs{B}$, we obtain
\begin{align}
 \delta f = \frac{(i\omega-i\bs{q}\cdot\bs{v})\bs{m}\cdot\bs{B}+\bs{v}\cdot e\bs{E}}{i\omega-i\bs{q}\cdot\bs{v}-1/\tau}\frac{\partial f(\ve)}{\partial\ve}.
\end{align}
The distribution function becomes
\begin{align}
 n
 &=f(\ve-\bs{m}\cdot\bs{B})+\frac{\omega-\bs{q}\cdot\bs{v}}{\omega-\bs{q}\cdot\bs{v}+i/\tau}(\bs{m}\cdot\bs{B})\frac{\partial f(\ve)}{\partial\ve}
 \nonumber\\
 &-\frac{i\bs{v}\cdot e\bs{E}}{\omega-\bs{q}\cdot\bs{v}+i/\tau}\frac{\partial f(\ve)}{\partial\ve}
 \nonumber\\
 &=f(\ve)-\frac{i/\tau}{\omega-\bs{q}\cdot\bs{v}+i/\tau}(\bs{m}\cdot\bs{B})\frac{\partial f(\ve)}{\partial\ve}
 \nonumber\\
 &-\frac{i\bs{v}\cdot e\bs{E}}{\omega-\bs{q}\cdot\bs{v}+i/\tau}\frac{\partial f(\ve)}{\partial\ve}.
\end{align}
Therefore, the current in the uniform limit reads~\cite{kharzeev2016anatomy}
\begin{align}
 \label{eq:dCME_decomp}
 \bs{J}=\bs{J}_{\mathrm{normal}}+\bs{J}_{\mathrm{CME}}+\bs{J}_{\mathrm{GME}}+\bs{J}_{\mathrm{mag}},
\end{align}
with
\begin{align}
 \bs{J}_{\mathrm{normal}}
 &= e\int_{\bs{k}}\bs{v}f(\ve),
 \label{eq:J_normal}
\\
 \bs{J}_{\mathrm{CME}}
 &=e^2\bs{B}\int_{\bs{k}}(\bs{v}\cdot\bs{\Omega})f(\ve)  ,
\label{eq:J_CME}
\\
 \bs{J}_{\mathrm{GME}}
 &=\frac{ie\omega\tau}{i\omega\tau-1}\int_{\bs{k}}\bs{v}(\bs{m}\cdot\bs{B})\frac{\partial f(\ve)}{\partial\ve},
\label{eq:J_GME}
\\
 \bs{J}_\mathrm{mag}
 &=\lim_{|\bs{q}|\to0}\frac{\partial}{\partial\bs{r}}\times\int_{\bs{k}}\bs{m}\frac{(1/\tau)\bs{m}\cdot\bs{B}+\bs{v}\cdot e\bs{E}}{i\omega-i\bs{q}\cdot\bs{v}-1/\tau}\frac{\partial f(\ve)}{\partial\ve}
 \nonumber\\
 &=\frac{e}{i\omega-1/\tau}\lim_{|\bs{q}|\to0}\frac{\partial}{\partial\bs{r}}\times\int_{\bs{k}}\bs{m}(\bs{v}\cdot\bs{E})\frac{\partial f(\ve)}{\partial\ve},
 \label{eq:J_mag}
\end{align}
We use a simplified notation $\int_{\bs{k}}\equiv\int\dif^3\bs{k}/(2\pi)^3$ only in this section. 
The first one is the normal current, and the sum of other three terms is the dynamic CME:
\be
\label{eq:dCME}
\bs{J}_{\mathrm{total}}=
\bs{J}_{\mathrm{CME}}+\bs{J}_{\mathrm{GME}}+\bs{J}_{\mathrm{mag}}. 
\ee
$\bs{J}_\mathrm{CME}$ is a usual static CME current, which survives even in the static limit $\omega\to0$.
$\bs{J}_\mathrm{GME}$ is the current induced by gyrotropic magnetic effect, which contributes to dynamic CME.
$\bs{J}_\mathrm{mag}$ also yields finite contribution to dynamic CME. 

Dynamic CME itself \textit{does not} show topological nature in general band structures because of $\bs{J}_{\mathrm{GME}}$ and $\bs{J}_{\mathrm{mag}}$. Only exception is the linear and isotropic band structure (See Appendix~\ref{app:2}). Another important property of the dynamic CME is that the gyrotropic magnetic and magnetization currents $\bs{J}_{\mathrm{GME}}$ and $\bs{J}_{\mathrm{mag}}$ can give finite contributions even when the chiral chemical potential vanishes. These two facts make a clear distinction of the dynamic CME from the static one.



\section{\label{sec:3}
CME in Multi-WSM}

In this section, we discuss the topological properties of the static and dynamic chiral magnetic conductivities of multi-WSMs in near-equilibrium states. The necessary formulas of the chiral kinetic theory are found in Sec.~\ref{app:1}.

\subsection{\label{sec:3a}
Static CME}

As we saw in the last section, the chiral magnetic current induced by a static magnetic field is provided by
\begin{align}
 \label{eq:CME0}
 \bs{J}_\mathrm{CME}&=-e^2\sum_{s=\pm}\int\frac{\dif^3\bs{k}}{(2\pi)^3}\left(\bs{\Omega}_{s}\cdot\frac{\partial f(\ve_s)}{\partial\bs{k}}\right)\ve_{s}\bs{B}
 \nonumber\\
 &\equiv \sigma_\mathrm{CME}\bs{B},
\end{align}
where $s=\pm$ refers to the contributions from the upper and lower Weyl cones. 
$\bs{\Omega}_{\pm}=(\Omega_{\pm}^x,\Omega_{\pm}^y,\Omega_{\pm}^z)$ are given by $\Omega_{\pm}^i={1\over 2}\epsilon^{ijk}\Omega_{\pm}^{jk}$ with $\Omega_{\pm}={1\over 2}\Omega_{\pm}^{ij}\dif k_i\wedge\dif k_j$, and $f(\ve)=(1+\mathrm{e}^{\beta(\ve-\mu)})^{-1}$ is the Fermi-Dirac distribution function. 
For the Weyl cones described by the model Hamiltonian \eqref{eq:mWH}, the chiral magnetic conductivity $\sigma_\mathrm{CME}$ at the zero temperature with positive doping is written as
\begin{align}
 \sigma_\mathrm{CME}
 =\frac{e^2\mu}{(2\pi)^3}\int\dif^{3}\bs{k} \delta(\ve_+(\bs{k})-\mu){\partial \ve_+(\bs{k})\over \partial\bs{k}}\cdot\bs{\Omega}_+. 
\label{eq:cme_conductivity}
\end{align}
The explicit and direct evaluation of the right-hand-side would require some tedious computation for $J>1$, so we here just mention its physical meaning and the result. We shall compute this quantity for more general situations in the moment.
$\int\dif^{3}\bs{k} \delta(\ve_+(\bs{k})-\mu)$ is an integration over the entire Fermi surface, and ${\partial \ve_+(\bs{k})\over \partial\bs{k}}$ is a vector normal to the Fermi surface. 
Since ${\partial \ve_+(\bs{k})\over \partial\bs{k}}\cdot \bs{\Omega}_+$ is the monopole flux penetrating the Fermi surface per unit area, the integration in Eq.~\eqref{eq:cme_conductivity} measures the total monopole flux penetrating the Fermi surface, which should be nothing but the monopole charge $K$ given in Eq.~\eqref{eq:monopole}. 
As a result, we should get $\bs{J}_\mathrm{CME}=(e^2/(2\pi)^2)K\mu\bs{B}$ for the static CME in a left-handed multi-Weyl fermion. 

In order to calculate the chiral magnetic conductivity for the general two-band Hamiltonian \eqref{eq:weyl_ham} at finite temperatures in an efficient manner, we rewrite $\sigma_{\mathrm{CME}}$ in Eq.~\eqref{eq:cme_conductivity}, by using differential forms, as (see Appendix~\ref{app:0})
\bea
\label{eq:CME1}
\sigma_{\mathrm{CME}}&=&-e^2\sum_{s=\pm}\int\frac{\dif^3\bs{k}}{(2\pi)^3}{\epsilon_{ijk}\over 2}{\Omega}_{s}^{jk}\frac{\partial f(\ve_s)}{\partial{k}_i}\ve_{s}\nonumber\\
&=&-{e^2\over (2\pi)^3}\sum_{s=\pm}\int \Omega_s \wedge \dif f(\ve_s) \ve_s. 
\eea
The energy eigenvalues of the Hamiltonian \eqref{eq:weyl_ham} is $\ve_{\pm}=\pm|\bs{R}|$, and the equilibrium distribution function $f(\ve_s(\bs{k}))$ depends on $\bs{k}$ only through $\pm|\bs{R}(\bs{k})|$. Also, $\Omega_{\pm}(\bs{k})$ depends only on $\hat{\bs{R}}$.  Then by using the expression~\eqref{eq:CME1}  the chiral magnetic conductivity can be readily computed as
\begin{align}
 \label{eq:general_CME}
 \sigma_\mathrm{CME}
 &=-\frac{e^2}{(2\pi)^3}\sum_{s=\pm}\int\Omega_s\wedge\dif f(\ve_s)\ve_s
 \nonumber\\
 &=-\frac{e^2}{(2\pi)^3}\int\ve_+\dif(f^+(\ve_+)-f^-(\ve_+)) \int \Omega_+
 \nonumber\\
 &=\frac{e^2}{(2\pi)^2}\mu K.
\end{align}
Here, we introduced the Fermi-Dirac distributions $f^+(\ve)=f(\ve)$ and $f^{-}(\ve)=1-f(-\ve)=(1+\mathrm{e}^{\beta(\ve+\mu)})^{-1}$ for the upper and lower Weyl cones, and we used the identity $\int r\dif r\frac{\partial}{\partial r}(f^+(r)-f^-(r))=-\mu$. 
Equation~\eqref{eq:general_CME} represents contribution only from the left-handed Weyl cones. Provided both of the left- and right-handed Weyl cones have the same monopole charge, the total chiral magnetic current is given by
\begin{align}
 \label{eq:CME_current}
 \bs{J}_{\mathrm{CME}}=\frac{e^2}{2\pi^2}K\mu_5\bs{B},
\end{align}
with $\mu_5\equiv(\mu_L-\mu_R)/2$. 
This extends the universal formula of the static CME in the linear WSMs to multi-WSMs by direct computation. 
We find that the chiral magnetic conductivity is independent of temperatures and the energy of Weyl points for not only the linearly dispersive WSMs but also multi-WSMs. 
Although Eq.~\eqref{eq:CME_current} was derived in Ref.~\cite{Son:2012wh} we do not assume that $\partial n/\partial \bs{k}=0$ at the monopole singularity, which was assumed in \cite{Son:2012wh}. This extension enables us to use Eq.~\eqref{eq:CME_current} in the case where Fermi level is close to monopole singularities. Furthermore, one can explicitly see that the chiral magnetic conductivity \eqref{eq:CME1} does not depend on metric in curved spacetime, meaning that it is robust against geometric deformations.

\subsection{Dynamic CME}
\label{sec:3b}

We calculate the dynamic chiral magnetic current \eqref{eq:dCME_decomp} for WSMs with general dispersion relations.
The dynamic CME is composed of the three contributions as in Eq.~\eqref{eq:dCME}.
The contribution from $\bs{J}_\mathrm{CME}$ was already calculated and given by Eq.~\eqref{eq:CME_current}.
We next discuss gyrotropic magnetic current $\bs{J}_\mathrm{GME}$ \eqref{eq:J_GME} and magnetization current $\bs{J}_\mathrm{mag}$ \eqref{eq:J_mag} by paying particular attention to their hidden topological properties. 

In the Weyl semimetals, since the spin degrees of freedom are decoupled, the magnetic moment $\bs{m}_s$ is given by the orbital magnetic moment~\cite{RevModPhys.82.1959},
\begin{align}
 \label{eq:orbit_mm}
 \bs{m}_s
 \equiv i\frac{e}{2}\left<\frac{\partial u_{s}}{\partial\bs{k}}\right|\times(H-\ve_{s})\left|\frac{\partial u_{s}}{\partial\bs{k}}\right>, 
\end{align}
Using differential forms, we can show that the magnetic moment $2$-form is related to the Berry curvature $2$-form as 
\begin{align}
 m_s
 &\equiv i\frac{e}{2}\left[\dif u^\dagger_{s}\wedge(H-\ve_s)\dif u_s\right]
 \nonumber\\
 &=i\frac{e}{2}(\ve_{-s}-\ve_s)\left[\dif u^\dagger_{s}\wedge(1-u_su^\dagger_s)\dif u_s\right]
 \nonumber\\
 &=-\frac{e}{2}(\ve_{-s}-\ve_s)\left[\dif\cA_s + i\cA_s\wedge\cA_s\right]
 \nonumber\\
 &=\frac{e}{2}(\ve_{s}-\ve_{-s})\Omega_s.
\end{align}
Here, we used the spectral decomposition of the Hamiltonian, $H=\sum_{s=\pm}\ve_s u_s u_s^{\dagger}$, and $1=\sum_{s=\pm}u_s u_s^{\dagger}$. 
Since we did not specify the energy dispersion relations, the relation between the magnetic moment and Berry curvature obtained here holds for general dispersion relations.

Although the GME and magnetization currents themselves are interesting, they are not topologically protected except for isotropic systems. Despite this fact, we now extract the topological nature hidden in those currents. First we consider the conductivity tensor of the GME (\ref{eq:J_GME}):
\begin{align}
 J_\mathrm{GME}^i=\sum_{j=x,y,z}\sigma_\mathrm{GME}^{ij}B_j.
\end{align}
By taking the trace of the conductivity tensor, we obtain
\begin{align}
 \label{eq:tr_GME}
 \sum_{i=x,y,z}\sigma_\mathrm{GME}^{ii}
 &=\frac{e}{(2\pi)^3}\sum_{s=\pm}\int m_s\wedge\dif f(\ve_s)
 \nonumber\\
 &=\frac{e^2}{(2\pi)^3}\sum_{s=\pm}\int \Omega_s\wedge\dif f(\ve_s)\frac{\ve_s-\ve_{-s}}{2},
\end{align}
which elucidates the topological property hidden in GME. This yields the same expression with Eq.~\eqref{eq:general_CME} for the general Weyl Hamiltonian \eqref{eq:weyl_ham}.

Next, we consider the magnetization current (\ref{eq:J_mag}) and extract the topological part of the conductivity.
Suppose we have the time-dependent external magnetic field in the $z$ direction, $\bs{B}=B\mathrm{e}^{-i\omega t+iqx}\hat{z}$. Then the corresponding electric field is $\bs{E}=B\mathrm{e}^{-i\omega t+iqx}\hat{y}$, which satisfies the Maxwell equation.  The induced current along the $z$ direction can be calculated as
\begin{align}
 J^z_\mathrm{mag}
 &=\frac{e}{i\omega-1/\tau}{\partial E_y\over \partial x}\sum_{s=\pm}\int\frac{\dif^3\bs{k}}{(2\pi)^3}m_{s}^y\frac{\partial f(\ve_s)}{\partial k_y}
 \nonumber\\
 &=\frac{e}{i\omega-1/\tau}(i\omega B_z)\sum_{s=\pm}\int\frac{\dif^3\bs{k}}{(2\pi)^3}m_{s}^y\frac{\partial f(\ve_s)}{\partial k_y}
 \nonumber\\
 &\equiv \sigma^z_\mathrm{mag}B_z.
\end{align}
The similar expression holds for the current in the $x$ or $y$ direction when a parallel external magnetic field is applied along that direction. Summing up the conductivities in these three cases, we obtain  the exactly same expression as that of the GME:
\begin{align}
 \label{eq:tr_mag}
 \sum_{i=x,y,z}\sigma_\mathrm{mag}^{i}
 =\frac{e^2}{(2\pi)^3}\sum_{s=\pm}\int \Omega_s\wedge\dif f(\ve_s)\frac{\ve_s-\ve_{-s}}{2},
\end{align}
in the clean limit.
We note that in general $\sigma_\mathrm{mag}^{x}\neq\sigma_\mathrm{mag}^{y}\neq\sigma_\mathrm{mag}^{z}$ in anisotropic systems.



\section{\label{sec:4}
Detection of monopole charge}

Here we discuss how to detect the monopole charge in multi-WSMs by experimental measurement of the chiral magnetic current.
Both static and dynamic CMEs yield the currents in the direction of the magnetic field at finite chiral chemical potential. However, the chiral chemical potential $\mu_5$ itself highly depends on the UV property of the band structure, which causes the chirality relaxation mechanism, even though Eq.~\eqref{eq:CME_current} is universal \cite{Fukushima:2008xe}. Hence, it is nontrivial how to extract the signal of the monopole charge from the static CME measurements. 

For $\mu_5=0$ the static CME clearly vanishes, but the dynamic CME is not necessarily zero. Let us think of the following Hamiltonian for the left- and right-handed Weyl fermions,
\begin{align}
 H_{L/R} = \pm \left(\sum_{i=1,2,3}R_i(\bs{k})\sigma_i +\frac{b_0}{2}\right),
\end{align}
where $\pm$ corresponds to the left- and right-handed Weyl fermions, respectively. The Hamiltonian describes a pair of Weyl cones whose Weyl points are shifted in energy by $b_0$.
Such a model Hamiltonian can be realized in WSMs with broken spatial-inversion symmetry~\cite{Ishizuka:2016dvm,Kharzeev:2016mvi}. 
The energy eigenvalues are $\ve_{L,\pm}=\pm|\bs{R}|+b_0/2$ and $\ve_{R,\pm}=\pm|\bs{R}|-b_0/2$. Therefore, for $\mu_5=0$ the trace part of the dynamic chiral magnetic conductivity for the total current consists of Eqs.~\eqref{eq:tr_GME} and \eqref{eq:tr_mag}:
\begin{align}
 \label{eq:tr_dCME}
 \sum_{i=1,2,3}\sigma_\mathrm{dCME}^{ii}
 =\frac{e^2}{2\pi^2}b_0 K.
\end{align}
We should emphasize here that this expression is robust against thermal excitations, deformation of Weyl cones and geometric deformation, which is certainly not true without taking trace. Furthermore, the quantity does not depend on UV property due to the absence of $\mu_5$ in contrast to the static CME as mentioned above.
Provided the energy shift of the Weyl points, the monopole charge can be extracted by measuring the dynamic chiral magnetic currents parallel to the external magnetic fields in the $x$, $y$, and $z$ directions and summing up each conductivity.

To extract the topological property from the observation of the chiral magnetic current, we need to require the absence of the anomalous Hall current, 
\be
\bs{J}_{\mathrm{AHE}}=\bs{E}\times\sum_{s=\pm}\int{\dif^3 \bs{k}\over (2\pi)^3}\bs{\Omega}_s f(\ve_s),
\label{eq:ahe}
\ee
which may give an unwanted contribution in the dynamical case for anisotropic systems. 
We raise three possibilities to eliminate the anomalous Hall contribution from measurements of the dynamic CME:
(i)~The anomalous Hall current vanishes in the presence of the time-reversal symmetry.
(ii)~The anomalous Hall conductivity vanishes under discrete rotational symmetries about two different directions. The point group symmetries ($C_4$ or $C_6$) protecting multi-WSMs can be important also for eliminating the anomalous Hall current. 
(iii)~External electric fields may be tuned to be small around the sample by using the standing electromagnetic wave.
An example of standing wave is $B_z=B_0\cos(q x)\cos(\omega t)$ and $E_y=E_0 \sin(q x)\sin(\omega t)$, then one can put the sample at $x=0$ when the sample size is negligible compared with $1/q$. 

A candidate material that fits to our proposal is SrSi$_2$, and it is numerically predicted to have the following three properties~\cite{huang2016new} (SrSi$_2$ was pointed out in previous works \cite{PhysRevB.92.205201,zhong2016gyrotropic,Kharzeev:2016mvi}): 
(A) The band structure possesses the Weyl points with double monopole charges.
(B) The crystal lacks both inversion and reflection symmetries, which makes it possible to observe the dynamic CME even without chiral chemical potential.
The numerical band computation indeed shows the finite energy shift between the left and right double-Weyl points ($b_0\sim1\times10^2\mathrm{[meV]}$).
(C) The system is time-reversal symmetric. Therefore, the anomalous Hall current should be excluded as mentioned above (i).


\section{\label{sec:5}
Conclusions}

We considered the general effective Hamiltonian describing multi-WSMs, and analyzed topological properties of the static and dynamic CMEs. 
In particular, the monopole charge of each Weyl point is characterized by the winding number of the map from $\bs{k}$ space to $\bs{R}$ space. 
We rewrite the expression of the static chiral magnetic conductivity using differential forms, which makes the topological feature of the static CME manifest. The obtained formula for multi-WSMs gives the straightforward extension of that for conventional WSMs with unit monopole charge. 
%
On the other hand, the dynamic CME in multi-WSMs is not manifestly topological, but we found the topological feature hidden there. 
We showed the general relation for multi-WSMs between the Berry curvature and orbital magnetic moment. Using the relation, the topological nature hidden in the dynamic CME is extracted by taking the trace of the gyrotropic magnetic and magnetization conductivities. 

We proposed an idea  to experimentally observe the multiple monopole charge through the measurement of the dynamic CME. 
The multi-WSMs must have the energy shift of a pair of Weyl points, i.e., the spatial inversion symmetry must be broken. 
In addition, the absence of reflection symmetry is necessary for the non-vanishing dynamic CME.
We also mentioned the anomalous Hall effect that can mask the signal of the dynamic CME, and several possibilities were discussed to evade the problem.
SrSi$_2$ is the WSMs satisfying all the criteria necessary for the detection of the multi-monopole charge.


\begin{acknowledgements}
The authors thank D. Kharzeev for reading the manuscript and making useful comments, particularly, on the experimental realization.
Authors also thank the organizers of the workshop ``Chiral Matter 2016'', where this work started.
Y.~K.~is supported by the Grants-in-Aid for JSPS fellows (Grant No.15J01626). 
T.~H.~is supported by JSPS Grant-in-Aid for Scientific Research (Grant No.~JP16J02240). 
Y.~T.~is supported by the Special Postdoctoral Researchers Program of RIKEN. 
\end{acknowledgements}


\appendix

\section{\label{app:0} 
Some formulas of differential forms}

We here briefly summarize some useful formulas, which are necessary for the calculations in the main text (see e.g., \cite{eguchi1980gravitation,nakahara2003geometry} for more details).
We only consider three dimensional flat space, which corresponds to $\bs{k}$ space in the main text. 

The differential line element $\dif x^i$ is called (differential) 1-form. The higher-forms are defined by antisymmetrizing the corresponding tensor fields. For instance, by using 1-forms $\dif x^{i,j}$ ($i,j=1,2,3$), the 2-form is defined in the following way:
\begin{align}
 \dif x^i\wedge\dif x^j &= \frac{1}{2}(\dif x^i\otimes\dif x^j-\dif x^j\otimes\dif x^i) 
 \nonumber
 \\
 &= -\dif x^j\wedge\dif x^i,
\end{align}
where $\otimes$ is the tensor product and $\wedge$ is called {\it wedge product}. The 3-form $\dif x^i\wedge\dif x^j\wedge\dif x^k$ is obtained by the same manner with antisymmetrizing the three indices. For $p$-form $\omega^p$ and $q$-form $\eta^q$, $\omega^p\wedge\eta^q=(-1)^{pq}\eta^q\wedge\omega^p$ is satisfied. Under the coordinate transformation $\tilde{x}^{a,b}=\tilde{x}^{a,b}(x),\ (a,b=1,2,3)$, the 2-form behaves as
\begin{align}
 \dif \tilde{x}^a\wedge\dif \tilde{x}^b
 = {1\over 2}\left(\frac{\partial\tilde{x}^a}{\partial x^i}\frac{\partial\tilde{x}^b}{\partial x^j}
 -\frac{\partial\tilde{x}^a}{\partial x^j}\frac{\partial\tilde{x}^b}{\partial x^i}\right)\dif x^i\wedge\dif x^j,
\end{align}
where the quantity inside the brackets is the Jacobian associated with the coordinate transformation.

The exterior derivative operation on differential forms, which turns $p$-forms into $(p+1)$-forms, is defined by
\begin{align}
 \dif \omega&=\frac{\partial\omega}{\partial x^i}\dif x^i,
 \\
 \dif (\omega_j\dif x^j)&=\frac{\partial\omega_j}{\partial x^i}\dif x^i\wedge\dif x^j,
 \\
 \dif (\omega_{jk}\dif x^j\wedge\dif x^k)&=\frac{\partial\omega_{jk}}{\partial x^i}\dif x^i\wedge\dif x^j\wedge\dif x^k,
\end{align}
where $\omega$, $\omega_j$, and $\omega_{jk}$ are scalar functions of $x$. 
The external derivative of a 3-form is identically zero in three dimensions.
The following property of the exterior derivative
\begin{align}
 \dif\dif\omega^p=0,
\end{align}
for an arbitrary $p$-form $\omega^p$ can be easily checked from the above definition.

The hodge star operation on differential forms in three-dimensional flat space, which turns $p$-forms into $(3-p)$-forms, is defined by
\begin{align}
 \ast\omega&=\frac{\epsilon_{ijk}}{3!}\omega\dif x^i\wedge\dif x^j\wedge\dif x^k,
 \\
 \ast(\omega_i\dif x^i)&=\frac{\epsilon_{ijk}}{2!}\omega_i\dif x^j\wedge\dif x^k,
 \\
 \ast(\omega_{ij}\dif x^i\wedge\dif x^j)&=\epsilon_{ijk}\omega_{ij}\dif x^k, 
 \\
 \ast(\omega_{ijk}\dif x^i\wedge\dif x^j\wedge\dif x^k)&=\epsilon_{ijk}\omega_{ijk}.
\end{align}

With the use of a differential p-form $\omega^p=\omega_{i_1\dots i_p}\dif x^{i_1}\wedge\dots\wedge\dif x^{i_p}$, the integration over the p-dimensional space $\mathcal{M}^p$ is defined by
\begin{align}
 \int_{\mathcal{M}^p}\omega^p=\int_{\mathcal{M}^p}\epsilon_{i_1\dots i_p}\omega_{i_1\dots i_p}\dif x^{1}\dots\dif x^{p},
\end{align}
where we used $\dif x^{i_1}\wedge\dots\wedge\dif x^{i_p}=\epsilon_{i_1\dots i_p}\dif x^{1}\wedge\dots\wedge\dif x^{p}$ and suppressed the wedge products in the right hand side. 
The integral of the inner product is expressed by using the hodge star operation. For instance, the integration of two 1-forms $\omega^1=\omega_i\dif x^i$ and $\eta^1=\eta_i\dif x^i$ over the three-dimensional space $\mathcal{M}^3$ is given as
\begin{align}
\label{eq:inner_pro}
 \int_{\mathcal{M}^3}\omega^1\wedge\ast\eta^1&=\int_{\mathcal{M}^3}(\omega_i\dif x^i)\wedge\left(\frac{\epsilon_{jkl}}{2}\eta_{j}\dif x^k\wedge\dif x^l\right)
 \nonumber\\
 &=\int_{\mathcal{M}^3}\omega_i\eta_{j}\frac{\epsilon_{jkl}}{2}\epsilon_{ikl}\dif x^1\dif x^2\dif x^3
 \nonumber\\
 &=\int_{\mathcal{M}^3}\omega_i\eta_i\dif x^1\dif x^2\dif x^3.
\end{align}

As an example, we see how the expression of CME in Eq.~\eqref{eq:CME0} is rewritten, by using the differential forms, as Eq.~\eqref{eq:CME1}.
First, the Berry connection 1-form $\cA$ and curvature 2-form $\Omega$ defined in Eq.~\eqref{eq:Berry_2form} are expressed as
\begin{align}
 \cA&=A^i\dif k_i,
 \\
 \Omega&=\dif\cA
\nonumber
 \\
 &=\frac{1}{2}\left(\frac{\partial A^j}{\partial k_i}-\frac{\partial A^i}{\partial k_j}\right)\dif k_i\wedge\dif k_j
 =\frac{1}{2}\Omega^{ij}\dif k_i\wedge\dif k_j .
\end{align}
It is noted that $\Omega^i=\frac{1}{2}\epsilon^{ijk}\Omega^{jk}$, in other words, the 1-form $\Omega^i\dif k_i$ is the hodge dual of the Berry curvature 2-form $\Omega={1\over 2}\Omega^{ij}\dif k_i\wedge\dif k_j$:
\begin{align}
 \Omega^i\dif k_i=\ast\left(\frac{1}{2}\Omega^{ij}\dif k_i\wedge\dif k_j\right)=\ast\Omega.
\end{align}
Similarly,
\begin{align}
 \label{eq:dual}
 \Omega =\ast(\Omega^i\dif k_i).
\end{align}
Therefore, the integration in Eq.~\eqref{eq:CME0} is written as
\begin{align}
\int \Omega^i_s\cdot\frac{\partial f}{\partial k_i}\ve_s\dif^3k
&=\int  \ast(\Omega^j_s\dif k_j) \wedge \ve_s\frac{\partial f}{\partial k_i}\dif k_i
\nonumber\\
&=\int \Omega_s \wedge \dif f \ve_s,
\end{align}
where we used Eq.~\eqref{eq:inner_pro} in the first equality and Eq.~\eqref{eq:dual} in the second equality.



\section{\label{app:2} 
Dynamic CME in isotropic two-band model}

One may wonder whether dynamic CME shows the topological properties as well as static CME, which is well known to be proportional to the monopole charge for isotropic and linear Hamiltonian.
It turned out that the chiral magnetic conductivity of dynamic CME has the topological property for an isotropic band as we will see below.
However, this is generally not true for anisotropic models such as multi-Weyl semimetals.
In order to figure out the difference between static and dynamic CME, we give the computation of dynamic CME for an isotropic two-band model with the linear dispersion relation.


%

If the system is isotropic, the gyrotropic magnetic current \eqref{eq:J_GME} and magnetization current \eqref{eq:J_mag} are further reduced to
\begin{align}
 \bs{J}_\mathrm{GME}
 &= \frac{ie\omega\tau}{i\omega\tau-1}\frac{1}{3}\bs{B}\int_{\bs{k}}\bs{m}\cdot\frac{\partial f(\ve)}{\partial\bs{k}},
 \\
  \bs{J}_\mathrm{mag}
 &=\frac{e}{i\omega-1/\tau}\frac{1}{3}\lim_{|\bs{q}|\to0}\frac{\partial}{\partial\bs{r}}\times\bs{E}\int_{\bs{k}}\bs{m}\cdot\frac{\partial f(\ve)}{\partial\bs{k}}
 \nonumber\\
 &=\frac{ie\omega\tau}{i\omega\tau-1}\frac{1}{3}\bs{B}\int_{\bs{k}}\bs{m}\cdot\frac{\partial f(\ve)}{\partial\bs{k}}
\end{align}
In the clean limit $\omega\tau\gg1$, we have
\begin{align}
 \bs{J}_\mathrm{GME}=\bs{J}_\mathrm{mag}=
  \frac{e}{3}\bs{B}\int_{\bs{k}}\bs{m}\cdot\frac{\partial f(\ve)}{\partial\bs{k}}.
\label{eq:J_GME_iso}
\end{align}
In this derivation, the isotropy of the GME or magnetization current is assumed, and thus the expression does not hold for anisotropic systems.

We apply the expressions given in Eqs.~\eqref{eq:J_CME} and~\eqref{eq:J_GME_iso} to the two-band model with linear band structure around each Weyl nodes for the purpose of illustrating the similarity and difference between the static and dynamic CMEs.
Consider the following Hamiltonian
\begin{align}
 H=\chi v_F\bs{k}\cdot\bs{\sigma},
\end{align}
near the Weyl point with the chirality $\chi=\pm1$.
In this case, the energy eigenvalues are $\ve_\pm=\pm \chi v_F |\bs{k}|$ corresponding to the upper and lower Weyl cones. Also, the Berry curvature is calculated to be $\bs{\Omega}_\pm=\pm\chi\bs{k}/(2|\bs{k}|^3)$. Hence, static chiral magnetic current takes the following form:
\begin{align}
 \bs{J}_\mathrm{CME}
 &= -e^2\bs{B}\int_{\bs{k},s}\left(\frac{s\chi\bs{k}}{2|\bs{k}|^3}\cdot\frac{\partial f(\ve_s)}{\partial\bs{k}}\right)\ve_s
\nonumber \\
& =\frac{e^2}{(2\pi)^2}\chi\mu\bs{B},
\end{align}
where we used the notation $\int_{\bs{k},s}\equiv\sum_{s=\pm}\int\dif^3\bs{k}/(2\pi)^3$ and $\varepsilon_s=s\chi v_F |\bs{k}|$ with $s=\pm1$.
Next we evaluate Eq.~\eqref{eq:J_GME_iso}. 

In this model, the orbital magnetic moment is calculated to be
\begin{align}
 \bs{m}_{\pm}=\bs{m}_{\pm}^\mathrm{orb}
 =\pm e\chi v_F\frac{\bs{k}}{2|\bs{k}|^2}.
\end{align}
Substituting it into Eq.~\eqref{eq:J_GME_iso}, we have
\begin{align}
 \bs{J}_\mathrm{GME}&=\bs{J}_\mathrm{mag}
 =\frac{e^2}{3}\bs{B}\int_{\bs{k},s}\frac{s\chi v_F\bs{k}}{2|\bs{k}|^2}\cdot\frac{\partial f(\ve_s)}{\partial\bs{k}}
 \nonumber\\
 &=-\frac{e^2}{3(2\pi)^2}\chi v_Fk_F\bs{B}.
\end{align}
Therefore, collecting the contributions to the dynamic CME, we obtain
\begin{align}
 \bs{J}_\mathrm{total}=\bs{J}_\mathrm{CME}+\bs{J}_\mathrm{GME}+\bs{J}_\mathrm{mag}
 \nonumber\\
 =\frac{e^2}{(2\pi)^2}\chi\left(\mu-\frac{2}{3}v_Fk_F\right)\bs{B}.
\end{align}
We should emphasize two points on this computation.
Firstly, the dynamic chiral magnetic conductivity is topological in the sense that it is proportional to the monopole charge $\chi$. This is because we assumed the isotropic band Hamiltonian, and not necessarily true for generic band structure. Nevertheless, we claim that the topological nature is hidden in \eqref{eq:J_GME} and \eqref{eq:J_mag} even in anisotropic cases as we will discuss in the main text.
Secondly, it is noted that the second term does not necessarily vanish even if the chiral chemical potential is zero ($\mu_5=(\mu_L-\mu_R)/2=0$).
Therefore, the mechanism of generating dynamic chiral magnetic current is indeed quite different from static one.
For instance, it is nonvanishing if locations of left- and right-handed Weyl nodes are shifted by finite energy~\cite{zhong2016gyrotropic}, which plays a crucial role in detecting monopole charge from dynamic CME measuremant.


\bibliography{references}

\begin{thebibliography}{45}%
\makeatletter
\providecommand \@ifxundefined [1]{%
 \@ifx{#1\undefined}
}%
\providecommand \@ifnum [1]{%
 \ifnum #1\expandafter \@firstoftwo
 \else \expandafter \@secondoftwo
 \fi
}%
\providecommand \@ifx [1]{%
 \ifx #1\expandafter \@firstoftwo
 \else \expandafter \@secondoftwo
 \fi
}%
\providecommand \natexlab [1]{#1}%
\providecommand \enquote  [1]{``#1''}%
\providecommand \bibnamefont  [1]{#1}%
\providecommand \bibfnamefont [1]{#1}%
\providecommand \citenamefont [1]{#1}%
\providecommand \href@noop [0]{\@secondoftwo}%
\providecommand \href [0]{\begingroup \@sanitize@url \@href}%
\providecommand \@href[1]{\@@startlink{#1}\@@href}%
\providecommand \@@href[1]{\endgroup#1\@@endlink}%
\providecommand \@sanitize@url [0]{\catcode `\\12\catcode `\$12\catcode
  `\&12\catcode `\#12\catcode `\^12\catcode `\_12\catcode `\%12\relax}%
\providecommand \@@startlink[1]{}%
\providecommand \@@endlink[0]{}%
\providecommand \url  [0]{\begingroup\@sanitize@url \@url }%
\providecommand \@url [1]{\endgroup\@href {#1}{\urlprefix }}%
\providecommand \urlprefix  [0]{URL }%
\providecommand \Eprint [0]{\href }%
\providecommand \doibase [0]{http://dx.doi.org/}%
\providecommand \selectlanguage [0]{\@gobble}%
\providecommand \bibinfo  [0]{\@secondoftwo}%
\providecommand \bibfield  [0]{\@secondoftwo}%
\providecommand \translation [1]{[#1]}%
\providecommand \BibitemOpen [0]{}%
\providecommand \bibitemStop [0]{}%
\providecommand \bibitemNoStop [0]{.\EOS\space}%
\providecommand \EOS [0]{\spacefactor3000\relax}%
\providecommand \BibitemShut  [1]{\csname bibitem#1\endcsname}%
\let\auto@bib@innerbib\@empty
\bibitem [{\citenamefont {{Murakami}}(2007)}]{Murakami}%
  \BibitemOpen
  \bibfield  {author} {\bibinfo {author} {\bibfnamefont {S.}~\bibnamefont
  {{Murakami}}},\ }\href {\doibase 10.1088/1367-2630/9/9/356} {\bibfield
  {journal} {\bibinfo  {journal} {New Journal of Physics}\ }\textbf {\bibinfo
  {volume} {9}},\ \bibinfo {pages} {356} (\bibinfo {year} {2007})}\BibitemShut
  {NoStop}%
\bibitem [{\citenamefont {Wan}\ \emph {et~al.}(2011)\citenamefont {Wan},
  \citenamefont {Turner}, \citenamefont {Vishwanath},\ and\ \citenamefont
  {Savrasov}}]{PhysRevB.83.205101}%
  \BibitemOpen
  \bibfield  {author} {\bibinfo {author} {\bibfnamefont {X.}~\bibnamefont
  {Wan}}, \bibinfo {author} {\bibfnamefont {A.~M.}\ \bibnamefont {Turner}},
  \bibinfo {author} {\bibfnamefont {A.}~\bibnamefont {Vishwanath}}, \ and\
  \bibinfo {author} {\bibfnamefont {S.~Y.}\ \bibnamefont {Savrasov}},\ }\href
  {\doibase 10.1103/PhysRevB.83.205101} {\bibfield  {journal} {\bibinfo
  {journal} {Phys. Rev. B}\ }\textbf {\bibinfo {volume} {83}},\ \bibinfo
  {pages} {205101} (\bibinfo {year} {2011})}\BibitemShut {NoStop}%
\bibitem [{\citenamefont {{Burkov}}\ and\ \citenamefont
  {{Balents}}(2011)}]{PhysRevLett.107.127205}%
  \BibitemOpen
  \bibfield  {author} {\bibinfo {author} {\bibfnamefont {A.~A.}\ \bibnamefont
  {{Burkov}}}\ and\ \bibinfo {author} {\bibfnamefont {L.}~\bibnamefont
  {{Balents}}},\ }\href {\doibase 10.1103/PhysRevLett.107.127205} {\bibfield
  {journal} {\bibinfo  {journal} {Phys. Rev. Lett.}\ }\textbf {\bibinfo
  {volume} {107}},\ \bibinfo {eid} {127205} (\bibinfo {year}
  {2011})}\BibitemShut {NoStop}%
\bibitem [{\citenamefont {Xu}\ \emph {et~al.}(2015)\citenamefont {Xu},
  \citenamefont {Belopolski}, \citenamefont {Alidoust}, \citenamefont
  {Neupane}, \citenamefont {Bian}, \citenamefont {Zhang}, \citenamefont
  {Sankar}, \citenamefont {Chang}, \citenamefont {Yuan}, \citenamefont {Lee},
  \citenamefont {Huang}, \citenamefont {Zheng}, \citenamefont {Ma},
  \citenamefont {Sanchez}, \citenamefont {Wang}, \citenamefont {Bansil},
  \citenamefont {Chou}, \citenamefont {Shibayev}, \citenamefont {Lin},
  \citenamefont {Jia},\ and\ \citenamefont {Hasan}}]{Xu613}%
  \BibitemOpen
  \bibfield  {author} {\bibinfo {author} {\bibfnamefont {S.-Y.}\ \bibnamefont
  {Xu}}, \bibinfo {author} {\bibfnamefont {I.}~\bibnamefont {Belopolski}},
  \bibinfo {author} {\bibfnamefont {N.}~\bibnamefont {Alidoust}}, \bibinfo
  {author} {\bibfnamefont {M.}~\bibnamefont {Neupane}}, \bibinfo {author}
  {\bibfnamefont {G.}~\bibnamefont {Bian}}, \bibinfo {author} {\bibfnamefont
  {C.}~\bibnamefont {Zhang}}, \bibinfo {author} {\bibfnamefont
  {R.}~\bibnamefont {Sankar}}, \bibinfo {author} {\bibfnamefont
  {G.}~\bibnamefont {Chang}}, \bibinfo {author} {\bibfnamefont
  {Z.}~\bibnamefont {Yuan}}, \bibinfo {author} {\bibfnamefont {C.-C.}\
  \bibnamefont {Lee}}, \bibinfo {author} {\bibfnamefont {S.-M.}\ \bibnamefont
  {Huang}}, \bibinfo {author} {\bibfnamefont {H.}~\bibnamefont {Zheng}},
  \bibinfo {author} {\bibfnamefont {J.}~\bibnamefont {Ma}}, \bibinfo {author}
  {\bibfnamefont {D.~S.}\ \bibnamefont {Sanchez}}, \bibinfo {author}
  {\bibfnamefont {B.}~\bibnamefont {Wang}}, \bibinfo {author} {\bibfnamefont
  {A.}~\bibnamefont {Bansil}}, \bibinfo {author} {\bibfnamefont
  {F.}~\bibnamefont {Chou}}, \bibinfo {author} {\bibfnamefont {P.~P.}\
  \bibnamefont {Shibayev}}, \bibinfo {author} {\bibfnamefont {H.}~\bibnamefont
  {Lin}}, \bibinfo {author} {\bibfnamefont {S.}~\bibnamefont {Jia}}, \ and\
  \bibinfo {author} {\bibfnamefont {M.~Z.}\ \bibnamefont {Hasan}},\ }\href
  {\doibase 10.1126/science.aaa9297} {\bibfield  {journal} {\bibinfo  {journal}
  {Science}\ }\textbf {\bibinfo {volume} {349}},\ \bibinfo {pages} {613}
  (\bibinfo {year} {2015})}\BibitemShut {NoStop}%
\bibitem [{\citenamefont {Lu}\ \emph {et~al.}(2015)\citenamefont {Lu},
  \citenamefont {Wang}, \citenamefont {Ye}, \citenamefont {Ran}, \citenamefont
  {Fu}, \citenamefont {Joannopoulos},\ and\ \citenamefont {Solja{\v
  c}i{\'c}}}]{Lu622}%
  \BibitemOpen
  \bibfield  {author} {\bibinfo {author} {\bibfnamefont {L.}~\bibnamefont
  {Lu}}, \bibinfo {author} {\bibfnamefont {Z.}~\bibnamefont {Wang}}, \bibinfo
  {author} {\bibfnamefont {D.}~\bibnamefont {Ye}}, \bibinfo {author}
  {\bibfnamefont {L.}~\bibnamefont {Ran}}, \bibinfo {author} {\bibfnamefont
  {L.}~\bibnamefont {Fu}}, \bibinfo {author} {\bibfnamefont {J.~D.}\
  \bibnamefont {Joannopoulos}}, \ and\ \bibinfo {author} {\bibfnamefont
  {M.}~\bibnamefont {Solja{\v c}i{\'c}}},\ }\href
  {http://science.sciencemag.org/content/349/6248/622} {\bibfield  {journal}
  {\bibinfo  {journal} {Science}\ }\textbf {\bibinfo {volume} {349}},\ \bibinfo
  {pages} {622} (\bibinfo {year} {2015})}\BibitemShut {NoStop}%
\bibitem [{\citenamefont {Lv}\ \emph {et~al.}(2015)\citenamefont {Lv},
  \citenamefont {Weng}, \citenamefont {Fu}, \citenamefont {Wang}, \citenamefont
  {Miao}, \citenamefont {Ma}, \citenamefont {Richard}, \citenamefont {Huang},
  \citenamefont {Zhao}, \citenamefont {Chen}, \citenamefont {Fang},
  \citenamefont {Dai}, \citenamefont {Qian},\ and\ \citenamefont
  {Ding}}]{PhysRevX.5.031013}%
  \BibitemOpen
  \bibfield  {author} {\bibinfo {author} {\bibfnamefont {B.~Q.}\ \bibnamefont
  {Lv}}, \bibinfo {author} {\bibfnamefont {H.~M.}\ \bibnamefont {Weng}},
  \bibinfo {author} {\bibfnamefont {B.~B.}\ \bibnamefont {Fu}}, \bibinfo
  {author} {\bibfnamefont {X.~P.}\ \bibnamefont {Wang}}, \bibinfo {author}
  {\bibfnamefont {H.}~\bibnamefont {Miao}}, \bibinfo {author} {\bibfnamefont
  {J.}~\bibnamefont {Ma}}, \bibinfo {author} {\bibfnamefont {P.}~\bibnamefont
  {Richard}}, \bibinfo {author} {\bibfnamefont {X.~C.}\ \bibnamefont {Huang}},
  \bibinfo {author} {\bibfnamefont {L.~X.}\ \bibnamefont {Zhao}}, \bibinfo
  {author} {\bibfnamefont {G.~F.}\ \bibnamefont {Chen}}, \bibinfo {author}
  {\bibfnamefont {Z.}~\bibnamefont {Fang}}, \bibinfo {author} {\bibfnamefont
  {X.}~\bibnamefont {Dai}}, \bibinfo {author} {\bibfnamefont {T.}~\bibnamefont
  {Qian}}, \ and\ \bibinfo {author} {\bibfnamefont {H.}~\bibnamefont {Ding}},\
  }\href {\doibase 10.1103/PhysRevX.5.031013} {\bibfield  {journal} {\bibinfo
  {journal} {Phys. Rev. X}\ }\textbf {\bibinfo {volume} {5}},\ \bibinfo {pages}
  {031013} (\bibinfo {year} {2015})}\BibitemShut {NoStop}%
\bibitem [{\citenamefont {Nielsen}\ and\ \citenamefont
  {Ninomiya}(1983)}]{Nielsen:1983rb}%
  \BibitemOpen
  \bibfield  {author} {\bibinfo {author} {\bibfnamefont {H.~B.}\ \bibnamefont
  {Nielsen}}\ and\ \bibinfo {author} {\bibfnamefont {M.}~\bibnamefont
  {Ninomiya}},\ }\href {\doibase 10.1016/0370-2693(83)91529-0} {\bibfield
  {journal} {\bibinfo  {journal} {Phys. Lett.}\ }\textbf {\bibinfo {volume}
  {B130}},\ \bibinfo {pages} {389} (\bibinfo {year} {1983})}\BibitemShut
  {NoStop}%
\bibitem [{\citenamefont {Fukushima}\ \emph {et~al.}(2008)\citenamefont
  {Fukushima}, \citenamefont {Kharzeev},\ and\ \citenamefont
  {Warringa}}]{Fukushima:2008xe}%
  \BibitemOpen
  \bibfield  {author} {\bibinfo {author} {\bibfnamefont {K.}~\bibnamefont
  {Fukushima}}, \bibinfo {author} {\bibfnamefont {D.~E.}\ \bibnamefont
  {Kharzeev}}, \ and\ \bibinfo {author} {\bibfnamefont {H.~J.}\ \bibnamefont
  {Warringa}},\ }\href {\doibase 10.1103/PhysRevD.78.074033} {\bibfield
  {journal} {\bibinfo  {journal} {Phys. Rev.}\ }\textbf {\bibinfo {volume}
  {D78}},\ \bibinfo {pages} {074033} (\bibinfo {year} {2008})}\BibitemShut
  {NoStop}%
\bibitem [{\citenamefont {{Son}}\ and\ \citenamefont
  {{Spivak}}(2013)}]{PhysRevB.88.104412}%
  \BibitemOpen
  \bibfield  {author} {\bibinfo {author} {\bibfnamefont {D.~T.}\ \bibnamefont
  {{Son}}}\ and\ \bibinfo {author} {\bibfnamefont {B.~Z.}\ \bibnamefont
  {{Spivak}}},\ }\href {\doibase 10.1103/PhysRevB.88.104412} {\bibfield
  {journal} {\bibinfo  {journal} {Phys. Rev.}\ }\textbf {\bibinfo {volume}
  {B88}},\ \bibinfo {eid} {104412} (\bibinfo {year} {2013})}\BibitemShut
  {NoStop}%
\bibitem [{\citenamefont {Li}\ \emph {et~al.}(2016)\citenamefont {Li},
  \citenamefont {Kharzeev}, \citenamefont {Zhang}, \citenamefont {Huang},
  \citenamefont {Pletikosic}, \citenamefont {Fedorov}, \citenamefont {Zhong},
  \citenamefont {Schneeloch}, \citenamefont {Gu},\ and\ \citenamefont
  {Valla}}]{KharzeevDirac}%
  \BibitemOpen
  \bibfield  {author} {\bibinfo {author} {\bibfnamefont {Q.}~\bibnamefont
  {Li}}, \bibinfo {author} {\bibfnamefont {D.~E.}\ \bibnamefont {Kharzeev}},
  \bibinfo {author} {\bibfnamefont {C.}~\bibnamefont {Zhang}}, \bibinfo
  {author} {\bibfnamefont {Y.}~\bibnamefont {Huang}}, \bibinfo {author}
  {\bibfnamefont {I.}~\bibnamefont {Pletikosic}}, \bibinfo {author}
  {\bibfnamefont {A.~V.}\ \bibnamefont {Fedorov}}, \bibinfo {author}
  {\bibfnamefont {R.~D.}\ \bibnamefont {Zhong}}, \bibinfo {author}
  {\bibfnamefont {J.~A.}\ \bibnamefont {Schneeloch}}, \bibinfo {author}
  {\bibfnamefont {G.~D.}\ \bibnamefont {Gu}}, \ and\ \bibinfo {author}
  {\bibfnamefont {T.}~\bibnamefont {Valla}},\ }\href
  {http://dx.doi.org/10.1038/nphys3648} {\bibfield  {journal} {\bibinfo
  {journal} {Nat. Phys.}\ }\textbf {\bibinfo {volume} {12}},\ \bibinfo {pages}
  {550} (\bibinfo {year} {2016})}\BibitemShut {NoStop}%
\bibitem [{\citenamefont {Xiong}\ \emph {et~al.}(2015)\citenamefont {Xiong},
  \citenamefont {Kushwaha}, \citenamefont {Liang}, \citenamefont {Krizan},
  \citenamefont {Hirschberger}, \citenamefont {Wang}, \citenamefont {Cava},\
  and\ \citenamefont {Ong}}]{xiong2015evidence}%
  \BibitemOpen
  \bibfield  {author} {\bibinfo {author} {\bibfnamefont {J.}~\bibnamefont
  {Xiong}}, \bibinfo {author} {\bibfnamefont {S.~K.}\ \bibnamefont {Kushwaha}},
  \bibinfo {author} {\bibfnamefont {T.}~\bibnamefont {Liang}}, \bibinfo
  {author} {\bibfnamefont {J.~W.}\ \bibnamefont {Krizan}}, \bibinfo {author}
  {\bibfnamefont {M.}~\bibnamefont {Hirschberger}}, \bibinfo {author}
  {\bibfnamefont {W.}~\bibnamefont {Wang}}, \bibinfo {author} {\bibfnamefont
  {R.}~\bibnamefont {Cava}}, \ and\ \bibinfo {author} {\bibfnamefont
  {N.}~\bibnamefont {Ong}},\ }\href {\doibase 10.1126/science.aac6089}
  {\bibfield  {journal} {\bibinfo  {journal} {Science}\ }\textbf {\bibinfo
  {volume} {350}},\ \bibinfo {pages} {413} (\bibinfo {year}
  {2015})}\BibitemShut {NoStop}%
\bibitem [{\citenamefont {Li}\ \emph {et~al.}(2015)\citenamefont {Li},
  \citenamefont {Wang}, \citenamefont {Liu}, \citenamefont {Wang},
  \citenamefont {Liao},\ and\ \citenamefont {Yu}}]{li2015giant}%
  \BibitemOpen
  \bibfield  {author} {\bibinfo {author} {\bibfnamefont {C.-Z.}\ \bibnamefont
  {Li}}, \bibinfo {author} {\bibfnamefont {L.-X.}\ \bibnamefont {Wang}},
  \bibinfo {author} {\bibfnamefont {H.}~\bibnamefont {Liu}}, \bibinfo {author}
  {\bibfnamefont {J.}~\bibnamefont {Wang}}, \bibinfo {author} {\bibfnamefont
  {Z.-M.}\ \bibnamefont {Liao}}, \ and\ \bibinfo {author} {\bibfnamefont
  {D.-P.}\ \bibnamefont {Yu}},\ }\href {\doibase 10.1038/ncomms10137}
  {\bibfield  {journal} {\bibinfo  {journal} {Nat. commun.}\ }\textbf {\bibinfo
  {volume} {6}},\ \bibinfo {pages} {10137} (\bibinfo {year}
  {2015})}\BibitemShut {NoStop}%
\bibitem [{\citenamefont {Huang}\ \emph {et~al.}(2015)\citenamefont {Huang},
  \citenamefont {Zhao}, \citenamefont {Long}, \citenamefont {Wang},
  \citenamefont {Chen}, \citenamefont {Yang}, \citenamefont {Liang},
  \citenamefont {Xue}, \citenamefont {Weng}, \citenamefont {Fang},
  \citenamefont {Dai},\ and\ \citenamefont {Chen}}]{huang2015observation}%
  \BibitemOpen
  \bibfield  {author} {\bibinfo {author} {\bibfnamefont {X.}~\bibnamefont
  {Huang}}, \bibinfo {author} {\bibfnamefont {L.}~\bibnamefont {Zhao}},
  \bibinfo {author} {\bibfnamefont {Y.}~\bibnamefont {Long}}, \bibinfo {author}
  {\bibfnamefont {P.}~\bibnamefont {Wang}}, \bibinfo {author} {\bibfnamefont
  {D.}~\bibnamefont {Chen}}, \bibinfo {author} {\bibfnamefont {Z.}~\bibnamefont
  {Yang}}, \bibinfo {author} {\bibfnamefont {H.}~\bibnamefont {Liang}},
  \bibinfo {author} {\bibfnamefont {M.}~\bibnamefont {Xue}}, \bibinfo {author}
  {\bibfnamefont {H.}~\bibnamefont {Weng}}, \bibinfo {author} {\bibfnamefont
  {Z.}~\bibnamefont {Fang}}, \bibinfo {author} {\bibfnamefont {X.}~\bibnamefont
  {Dai}}, \ and\ \bibinfo {author} {\bibfnamefont {G.}~\bibnamefont {Chen}},\
  }\href {\doibase 10.1103/PhysRevX.5.031023} {\bibfield  {journal} {\bibinfo
  {journal} {Phys. Rev. X}\ }\textbf {\bibinfo {volume} {5}},\ \bibinfo {pages}
  {031023} (\bibinfo {year} {2015})}\BibitemShut {NoStop}%
\bibitem [{\citenamefont {Wang}\ \emph {et~al.}(2016)\citenamefont {Wang},
  \citenamefont {Zheng}, \citenamefont {Shen}, \citenamefont {Lu},
  \citenamefont {Fang}, \citenamefont {Sheng}, \citenamefont {Zhou},
  \citenamefont {Yang}, \citenamefont {Li}, \citenamefont {Feng},\ and\
  \citenamefont {Xu}}]{wang2015helicity}%
  \BibitemOpen
  \bibfield  {author} {\bibinfo {author} {\bibfnamefont {Z.}~\bibnamefont
  {Wang}}, \bibinfo {author} {\bibfnamefont {Y.}~\bibnamefont {Zheng}},
  \bibinfo {author} {\bibfnamefont {Z.}~\bibnamefont {Shen}}, \bibinfo {author}
  {\bibfnamefont {Y.}~\bibnamefont {Lu}}, \bibinfo {author} {\bibfnamefont
  {H.}~\bibnamefont {Fang}}, \bibinfo {author} {\bibfnamefont {F.}~\bibnamefont
  {Sheng}}, \bibinfo {author} {\bibfnamefont {Y.}~\bibnamefont {Zhou}},
  \bibinfo {author} {\bibfnamefont {X.}~\bibnamefont {Yang}}, \bibinfo {author}
  {\bibfnamefont {Y.}~\bibnamefont {Li}}, \bibinfo {author} {\bibfnamefont
  {C.}~\bibnamefont {Feng}}, \ and\ \bibinfo {author} {\bibfnamefont {Z.-A.}\
  \bibnamefont {Xu}},\ }\href {\doibase 10.1103/PhysRevB.93.121112} {\bibfield
  {journal} {\bibinfo  {journal} {Phys. Rev. B}\ }\textbf {\bibinfo {volume}
  {93}},\ \bibinfo {pages} {121112} (\bibinfo {year} {2016})}\BibitemShut
  {NoStop}%
\bibitem [{\citenamefont {Zhang}\ \emph {et~al.}(2016)\citenamefont {Zhang},
  \citenamefont {Xu}, \citenamefont {Belopolski}, \citenamefont {Yuan},
  \citenamefont {Lin}, \citenamefont {Tong}, \citenamefont {Bian},
  \citenamefont {Alidoust}, \citenamefont {Lee}, \citenamefont {Huang},
  \citenamefont {Chang}, \citenamefont {Chang}, \citenamefont {Hsu},
  \citenamefont {Jeng}, \citenamefont {Neupane}, \citenamefont {Sanchez},
  \citenamefont {Zheng}, \citenamefont {Wang}, \citenamefont {Lin},
  \citenamefont {Zhang}, \citenamefont {Lu}, \citenamefont {Shen},
  \citenamefont {Neupert}, \citenamefont {Zahid~Hasan},\ and\ \citenamefont
  {Jia}}]{zhang2015observation}%
  \BibitemOpen
  \bibfield  {author} {\bibinfo {author} {\bibfnamefont {C.-L.}\ \bibnamefont
  {Zhang}}, \bibinfo {author} {\bibfnamefont {S.-Y.}\ \bibnamefont {Xu}},
  \bibinfo {author} {\bibfnamefont {I.}~\bibnamefont {Belopolski}}, \bibinfo
  {author} {\bibfnamefont {Z.}~\bibnamefont {Yuan}}, \bibinfo {author}
  {\bibfnamefont {Z.}~\bibnamefont {Lin}}, \bibinfo {author} {\bibfnamefont
  {B.}~\bibnamefont {Tong}}, \bibinfo {author} {\bibfnamefont {G.}~\bibnamefont
  {Bian}}, \bibinfo {author} {\bibfnamefont {N.}~\bibnamefont {Alidoust}},
  \bibinfo {author} {\bibfnamefont {C.-C.}\ \bibnamefont {Lee}}, \bibinfo
  {author} {\bibfnamefont {S.-M.}\ \bibnamefont {Huang}}, \bibinfo {author}
  {\bibfnamefont {T.-R.}\ \bibnamefont {Chang}}, \bibinfo {author}
  {\bibfnamefont {G.}~\bibnamefont {Chang}}, \bibinfo {author} {\bibfnamefont
  {C.-H.}\ \bibnamefont {Hsu}}, \bibinfo {author} {\bibfnamefont {H.-T.}\
  \bibnamefont {Jeng}}, \bibinfo {author} {\bibfnamefont {M.}~\bibnamefont
  {Neupane}}, \bibinfo {author} {\bibfnamefont {D.~S.}\ \bibnamefont
  {Sanchez}}, \bibinfo {author} {\bibfnamefont {H.}~\bibnamefont {Zheng}},
  \bibinfo {author} {\bibfnamefont {J.}~\bibnamefont {Wang}}, \bibinfo {author}
  {\bibfnamefont {H.}~\bibnamefont {Lin}}, \bibinfo {author} {\bibfnamefont
  {C.}~\bibnamefont {Zhang}}, \bibinfo {author} {\bibfnamefont {H.-Z.}\
  \bibnamefont {Lu}}, \bibinfo {author} {\bibfnamefont {S.-Q.}\ \bibnamefont
  {Shen}}, \bibinfo {author} {\bibfnamefont {T.}~\bibnamefont {Neupert}},
  \bibinfo {author} {\bibfnamefont {M.}~\bibnamefont {Zahid~Hasan}}, \ and\
  \bibinfo {author} {\bibfnamefont {S.}~\bibnamefont {Jia}},\ }\href
  {http://dx.doi.org/10.1038/ncomms10735} {\bibfield  {journal} {\bibinfo
  {journal} {Nat. Commun.}\ }\textbf {\bibinfo {volume} {7}},\ \bibinfo {pages}
  {10735} (\bibinfo {year} {2016})}\BibitemShut {NoStop}%
\bibitem [{\citenamefont {Arnold}\ \emph {et~al.}(2016)\citenamefont {Arnold},
  \citenamefont {Shekhar}, \citenamefont {Wu}, \citenamefont {Sun},
  \citenamefont {dos Reis}, \citenamefont {Kumar}, \citenamefont {Naumann},
  \citenamefont {Ajeesh}, \citenamefont {Schmidt}, \citenamefont {Grushin},
  \citenamefont {Bardarson}, \citenamefont {Baenitz}, \citenamefont {Sokolov},
  \citenamefont {Borrmann}, \citenamefont {Nicklas}, \citenamefont {Felser},
  \citenamefont {Hassinger},\ and\ \citenamefont {Yan}}]{shekhar2015large}%
  \BibitemOpen
  \bibfield  {author} {\bibinfo {author} {\bibfnamefont {F.}~\bibnamefont
  {Arnold}}, \bibinfo {author} {\bibfnamefont {C.}~\bibnamefont {Shekhar}},
  \bibinfo {author} {\bibfnamefont {S.-C.}\ \bibnamefont {Wu}}, \bibinfo
  {author} {\bibfnamefont {Y.}~\bibnamefont {Sun}}, \bibinfo {author}
  {\bibfnamefont {R.~D.}\ \bibnamefont {dos Reis}}, \bibinfo {author}
  {\bibfnamefont {N.}~\bibnamefont {Kumar}}, \bibinfo {author} {\bibfnamefont
  {M.}~\bibnamefont {Naumann}}, \bibinfo {author} {\bibfnamefont {M.~O.}\
  \bibnamefont {Ajeesh}}, \bibinfo {author} {\bibfnamefont {M.}~\bibnamefont
  {Schmidt}}, \bibinfo {author} {\bibfnamefont {A.~G.}\ \bibnamefont
  {Grushin}}, \bibinfo {author} {\bibfnamefont {J.~H.}\ \bibnamefont
  {Bardarson}}, \bibinfo {author} {\bibfnamefont {M.}~\bibnamefont {Baenitz}},
  \bibinfo {author} {\bibfnamefont {D.}~\bibnamefont {Sokolov}}, \bibinfo
  {author} {\bibfnamefont {H.}~\bibnamefont {Borrmann}}, \bibinfo {author}
  {\bibfnamefont {M.}~\bibnamefont {Nicklas}}, \bibinfo {author} {\bibfnamefont
  {C.}~\bibnamefont {Felser}}, \bibinfo {author} {\bibfnamefont
  {E.}~\bibnamefont {Hassinger}}, \ and\ \bibinfo {author} {\bibfnamefont
  {B.}~\bibnamefont {Yan}},\ }\href {\doibase 10.1038/ncomms11615} {\bibfield
  {journal} {\bibinfo  {journal} {Nat. Commun.}\ }\textbf {\bibinfo {volume}
  {7}},\ \bibinfo {pages} {11615} (\bibinfo {year} {2016})}\BibitemShut
  {NoStop}%
\bibitem [{\citenamefont {Xu}\ \emph {et~al.}(2011)\citenamefont {Xu},
  \citenamefont {Weng}, \citenamefont {Wang}, \citenamefont {Dai},\ and\
  \citenamefont {Fang}}]{PhysRevLett.107.186806}%
  \BibitemOpen
  \bibfield  {author} {\bibinfo {author} {\bibfnamefont {G.}~\bibnamefont
  {Xu}}, \bibinfo {author} {\bibfnamefont {H.}~\bibnamefont {Weng}}, \bibinfo
  {author} {\bibfnamefont {Z.}~\bibnamefont {Wang}}, \bibinfo {author}
  {\bibfnamefont {X.}~\bibnamefont {Dai}}, \ and\ \bibinfo {author}
  {\bibfnamefont {Z.}~\bibnamefont {Fang}},\ }\href {\doibase
  10.1103/PhysRevLett.107.186806} {\bibfield  {journal} {\bibinfo  {journal}
  {Phys. Rev. Lett.}\ }\textbf {\bibinfo {volume} {107}},\ \bibinfo {pages}
  {186806} (\bibinfo {year} {2011})}\BibitemShut {NoStop}%
\bibitem [{\citenamefont {Fang}\ \emph {et~al.}(2012)\citenamefont {Fang},
  \citenamefont {Gilbert}, \citenamefont {Dai},\ and\ \citenamefont
  {Bernevig}}]{fang2012multi}%
  \BibitemOpen
  \bibfield  {author} {\bibinfo {author} {\bibfnamefont {C.}~\bibnamefont
  {Fang}}, \bibinfo {author} {\bibfnamefont {M.~J.}\ \bibnamefont {Gilbert}},
  \bibinfo {author} {\bibfnamefont {X.}~\bibnamefont {Dai}}, \ and\ \bibinfo
  {author} {\bibfnamefont {B.~A.}\ \bibnamefont {Bernevig}},\ }\href {\doibase
  10.1103/PhysRevLett.108.266802} {\bibfield  {journal} {\bibinfo  {journal}
  {Phys. Rev. Lett.}\ }\textbf {\bibinfo {volume} {108}},\ \bibinfo {pages}
  {266802} (\bibinfo {year} {2012})}\BibitemShut {NoStop}%
\bibitem [{\citenamefont {Chen}\ and\ \citenamefont
  {Fiete}(2016)}]{PhysRevB.93.155125}%
  \BibitemOpen
  \bibfield  {author} {\bibinfo {author} {\bibfnamefont {Q.}~\bibnamefont
  {Chen}}\ and\ \bibinfo {author} {\bibfnamefont {G.~A.}\ \bibnamefont
  {Fiete}},\ }\href {\doibase 10.1103/PhysRevB.93.155125} {\bibfield  {journal}
  {\bibinfo  {journal} {Phys. Rev. B}\ }\textbf {\bibinfo {volume} {93}},\
  \bibinfo {pages} {155125} (\bibinfo {year} {2016})}\BibitemShut {NoStop}%
\bibitem [{\citenamefont {{Park}}\ \emph {et~al.}()\citenamefont {{Park}},
  \citenamefont {{Woo}}, \citenamefont {{Mele}},\ and\ \citenamefont
  {{Min}}}]{park2017semiclassical}%
  \BibitemOpen
  \bibfield  {author} {\bibinfo {author} {\bibfnamefont {S.}~\bibnamefont
  {{Park}}}, \bibinfo {author} {\bibfnamefont {S.}~\bibnamefont {{Woo}}},
  \bibinfo {author} {\bibfnamefont {E.~J.}\ \bibnamefont {{Mele}}}, \ and\
  \bibinfo {author} {\bibfnamefont {H.}~\bibnamefont {{Min}}},\ }\href@noop {}
  {\ }\Eprint {http://arxiv.org/abs/1701.07578} {arXiv:1701.07578
  [cond-mat.mes-hall]} \BibitemShut {NoStop}%
\bibitem [{\citenamefont {Vazifeh}\ and\ \citenamefont
  {Franz}(2013)}]{PhysRevLett.111.027201}%
  \BibitemOpen
  \bibfield  {author} {\bibinfo {author} {\bibfnamefont {M.~M.}\ \bibnamefont
  {Vazifeh}}\ and\ \bibinfo {author} {\bibfnamefont {M.}~\bibnamefont
  {Franz}},\ }\href {\doibase 10.1103/PhysRevLett.111.027201} {\bibfield
  {journal} {\bibinfo  {journal} {Phys. Rev. Lett.}\ }\textbf {\bibinfo
  {volume} {111}},\ \bibinfo {pages} {027201} (\bibinfo {year}
  {2013})}\BibitemShut {NoStop}%
\bibitem [{\citenamefont {Chen}\ \emph
  {et~al.}(2013{\natexlab{a}})\citenamefont {Chen}, \citenamefont {Wu},\ and\
  \citenamefont {Burkov}}]{PhysRevB.88.125105}%
  \BibitemOpen
  \bibfield  {author} {\bibinfo {author} {\bibfnamefont {Y.}~\bibnamefont
  {Chen}}, \bibinfo {author} {\bibfnamefont {S.}~\bibnamefont {Wu}}, \ and\
  \bibinfo {author} {\bibfnamefont {A.~A.}\ \bibnamefont {Burkov}},\ }\href
  {\doibase 10.1103/PhysRevB.88.125105} {\bibfield  {journal} {\bibinfo
  {journal} {Phys. Rev. B}\ }\textbf {\bibinfo {volume} {88}},\ \bibinfo
  {pages} {125105} (\bibinfo {year} {2013}{\natexlab{a}})}\BibitemShut
  {NoStop}%
\bibitem [{\citenamefont {Basar}\ \emph {et~al.}(2014)\citenamefont {Basar},
  \citenamefont {Kharzeev},\ and\ \citenamefont {Yee}}]{Basar:2013iaa}%
  \BibitemOpen
  \bibfield  {author} {\bibinfo {author} {\bibfnamefont {G.}~\bibnamefont
  {Basar}}, \bibinfo {author} {\bibfnamefont {D.~E.}\ \bibnamefont {Kharzeev}},
  \ and\ \bibinfo {author} {\bibfnamefont {H.-U.}\ \bibnamefont {Yee}},\ }\href
  {\doibase 10.1103/PhysRevB.89.035142} {\bibfield  {journal} {\bibinfo
  {journal} {Phys. Rev.}\ }\textbf {\bibinfo {volume} {B89}},\ \bibinfo {pages}
  {035142} (\bibinfo {year} {2014})},\ \Eprint {http://arxiv.org/abs/1305.6338}
  {arXiv:1305.6338 [hep-th]} \BibitemShut {NoStop}%
\bibitem [{\citenamefont {Landsteiner}(2014)}]{PhysRevB.89.075124}%
  \BibitemOpen
  \bibfield  {author} {\bibinfo {author} {\bibfnamefont {K.}~\bibnamefont
  {Landsteiner}},\ }\href {\doibase 10.1103/PhysRevB.89.075124} {\bibfield
  {journal} {\bibinfo  {journal} {Phys. Rev. B}\ }\textbf {\bibinfo {volume}
  {89}},\ \bibinfo {pages} {075124} (\bibinfo {year} {2014})}\BibitemShut
  {NoStop}%
\bibitem [{\citenamefont {Yamamoto}(2015)}]{PhysRevD.92.085011}%
  \BibitemOpen
  \bibfield  {author} {\bibinfo {author} {\bibfnamefont {N.}~\bibnamefont
  {Yamamoto}},\ }\href {\doibase 10.1103/PhysRevD.92.085011} {\bibfield
  {journal} {\bibinfo  {journal} {Phys. Rev. D}\ }\textbf {\bibinfo {volume}
  {92}},\ \bibinfo {pages} {085011} (\bibinfo {year} {2015})}\BibitemShut
  {NoStop}%
\bibitem [{\citenamefont {Chang}\ and\ \citenamefont
  {Yang}(2015{\natexlab{a}})}]{PhysRevB.91.115203}%
  \BibitemOpen
  \bibfield  {author} {\bibinfo {author} {\bibfnamefont {M.-C.}\ \bibnamefont
  {Chang}}\ and\ \bibinfo {author} {\bibfnamefont {M.-F.}\ \bibnamefont
  {Yang}},\ }\href {\doibase 10.1103/PhysRevB.91.115203} {\bibfield  {journal}
  {\bibinfo  {journal} {Phys. Rev. B}\ }\textbf {\bibinfo {volume} {91}},\
  \bibinfo {pages} {115203} (\bibinfo {year} {2015}{\natexlab{a}})}\BibitemShut
  {NoStop}%
\bibitem [{\citenamefont {Xiao}\ \emph {et~al.}(2010)\citenamefont {Xiao},
  \citenamefont {Chang},\ and\ \citenamefont {Niu}}]{RevModPhys.82.1959}%
  \BibitemOpen
  \bibfield  {author} {\bibinfo {author} {\bibfnamefont {D.}~\bibnamefont
  {Xiao}}, \bibinfo {author} {\bibfnamefont {M.-C.}\ \bibnamefont {Chang}}, \
  and\ \bibinfo {author} {\bibfnamefont {Q.}~\bibnamefont {Niu}},\ }\href
  {\doibase 10.1103/RevModPhys.82.1959} {\bibfield  {journal} {\bibinfo
  {journal} {Rev. Mod. Phys.}\ }\textbf {\bibinfo {volume} {82}},\ \bibinfo
  {pages} {1959} (\bibinfo {year} {2010})}\BibitemShut {NoStop}%
\bibitem [{\citenamefont {Son}\ and\ \citenamefont
  {Yamamoto}(2013)}]{PhysRevD.87.085016}%
  \BibitemOpen
  \bibfield  {author} {\bibinfo {author} {\bibfnamefont {D.~T.}\ \bibnamefont
  {Son}}\ and\ \bibinfo {author} {\bibfnamefont {N.}~\bibnamefont {Yamamoto}},\
  }\href {\doibase 10.1103/PhysRevD.87.085016} {\bibfield  {journal} {\bibinfo
  {journal} {Phys. Rev. D}\ }\textbf {\bibinfo {volume} {87}},\ \bibinfo
  {pages} {085016} (\bibinfo {year} {2013})}\BibitemShut {NoStop}%
\bibitem [{\citenamefont {Alavirad}\ and\ \citenamefont
  {Sau}(2016)}]{PhysRevB.94.115160}%
  \BibitemOpen
  \bibfield  {author} {\bibinfo {author} {\bibfnamefont {Y.}~\bibnamefont
  {Alavirad}}\ and\ \bibinfo {author} {\bibfnamefont {J.~D.}\ \bibnamefont
  {Sau}},\ }\href {\doibase 10.1103/PhysRevB.94.115160} {\bibfield  {journal}
  {\bibinfo  {journal} {Phys. Rev. B}\ }\textbf {\bibinfo {volume} {94}},\
  \bibinfo {pages} {115160} (\bibinfo {year} {2016})}\BibitemShut {NoStop}%
\bibitem [{\citenamefont {Zhong}\ \emph {et~al.}(2016)\citenamefont {Zhong},
  \citenamefont {Moore},\ and\ \citenamefont {Souza}}]{zhong2016gyrotropic}%
  \BibitemOpen
  \bibfield  {author} {\bibinfo {author} {\bibfnamefont {S.}~\bibnamefont
  {Zhong}}, \bibinfo {author} {\bibfnamefont {J.~E.}\ \bibnamefont {Moore}}, \
  and\ \bibinfo {author} {\bibfnamefont {I.}~\bibnamefont {Souza}},\ }\href
  {\doibase 10.1103/PhysRevLett.116.077201} {\bibfield  {journal} {\bibinfo
  {journal} {Phys. Rev. Lett.}\ }\textbf {\bibinfo {volume} {116}},\ \bibinfo
  {pages} {077201} (\bibinfo {year} {2016})}\BibitemShut {NoStop}%
\bibitem [{\citenamefont {{Kharzeev}}\ \emph {et~al.}()\citenamefont
  {{Kharzeev}}, \citenamefont {{Stephanov}},\ and\ \citenamefont
  {{Yee}}}]{kharzeev2016anatomy}%
  \BibitemOpen
  \bibfield  {author} {\bibinfo {author} {\bibfnamefont {D.~E.}\ \bibnamefont
  {{Kharzeev}}}, \bibinfo {author} {\bibfnamefont {M.~A.}\ \bibnamefont
  {{Stephanov}}}, \ and\ \bibinfo {author} {\bibfnamefont {H.-U.}\ \bibnamefont
  {{Yee}}},\ }\href@noop {} {\ }\Eprint {http://arxiv.org/abs/1612.01674}
  {arXiv:1612.01674 [hep-ph]} \BibitemShut {NoStop}%
\bibitem [{\citenamefont {Ma}\ and\ \citenamefont
  {Pesin}(2017)}]{PhysRevLett.118.107401}%
  \BibitemOpen
  \bibfield  {author} {\bibinfo {author} {\bibfnamefont {J.}~\bibnamefont
  {Ma}}\ and\ \bibinfo {author} {\bibfnamefont {D.~A.}\ \bibnamefont {Pesin}},\
  }\href {\doibase 10.1103/PhysRevLett.118.107401} {\bibfield  {journal}
  {\bibinfo  {journal} {Phys. Rev. Lett.}\ }\textbf {\bibinfo {volume} {118}},\
  \bibinfo {pages} {107401} (\bibinfo {year} {2017})}\BibitemShut {NoStop}%
\bibitem [{\citenamefont {Stephanov}\ and\ \citenamefont
  {Yin}(2012)}]{Stephanov:2012ki}%
  \BibitemOpen
  \bibfield  {author} {\bibinfo {author} {\bibfnamefont {M.~A.}\ \bibnamefont
  {Stephanov}}\ and\ \bibinfo {author} {\bibfnamefont {Y.}~\bibnamefont
  {Yin}},\ }\href {\doibase 10.1103/PhysRevLett.109.162001} {\bibfield
  {journal} {\bibinfo  {journal} {Phys. Rev. Lett.}\ }\textbf {\bibinfo
  {volume} {109}},\ \bibinfo {pages} {162001} (\bibinfo {year}
  {2012})}\BibitemShut {NoStop}%
\bibitem [{\citenamefont {Son}\ and\ \citenamefont
  {Yamamoto}(2012)}]{Son:2012wh}%
  \BibitemOpen
  \bibfield  {author} {\bibinfo {author} {\bibfnamefont {D.~T.}\ \bibnamefont
  {Son}}\ and\ \bibinfo {author} {\bibfnamefont {N.}~\bibnamefont {Yamamoto}},\
  }\href {\doibase 10.1103/PhysRevLett.109.181602} {\bibfield  {journal}
  {\bibinfo  {journal} {Phys. Rev. Lett.}\ }\textbf {\bibinfo {volume} {109}},\
  \bibinfo {pages} {181602} (\bibinfo {year} {2012})}\BibitemShut {NoStop}%
\bibitem [{\citenamefont {Chen}\ \emph
  {et~al.}(2013{\natexlab{b}})\citenamefont {Chen}, \citenamefont {Pu},
  \citenamefont {Wang},\ and\ \citenamefont {Wang}}]{Chen:2012ca}%
  \BibitemOpen
  \bibfield  {author} {\bibinfo {author} {\bibfnamefont {J.-W.}\ \bibnamefont
  {Chen}}, \bibinfo {author} {\bibfnamefont {S.}~\bibnamefont {Pu}}, \bibinfo
  {author} {\bibfnamefont {Q.}~\bibnamefont {Wang}}, \ and\ \bibinfo {author}
  {\bibfnamefont {X.-N.}\ \bibnamefont {Wang}},\ }\href {\doibase
  10.1103/PhysRevLett.110.262301} {\bibfield  {journal} {\bibinfo  {journal}
  {Phys. Rev. Lett.}\ }\textbf {\bibinfo {volume} {110}},\ \bibinfo {pages}
  {262301} (\bibinfo {year} {2013}{\natexlab{b}})}\BibitemShut {NoStop}%
\bibitem [{\citenamefont {Sundaram}\ and\ \citenamefont
  {Niu}(1999)}]{PhysRevB.59.14915}%
  \BibitemOpen
  \bibfield  {author} {\bibinfo {author} {\bibfnamefont {G.}~\bibnamefont
  {Sundaram}}\ and\ \bibinfo {author} {\bibfnamefont {Q.}~\bibnamefont {Niu}},\
  }\href {\doibase 10.1103/PhysRevB.59.14915} {\bibfield  {journal} {\bibinfo
  {journal} {Phys. Rev. B}\ }\textbf {\bibinfo {volume} {59}},\ \bibinfo
  {pages} {14915} (\bibinfo {year} {1999})}\BibitemShut {NoStop}%
\bibitem [{Note1()}]{Note1}%
  \BibitemOpen
  \bibinfo {note} {In this paper, we define the chiral chemical potential $\mu
  _5$ in such a way that $\mu _5$ vanishes in the equilibrium state even if
  there is a energy separation $b_0$ between left and right handed Weyl
  cones.}\BibitemShut {Stop}%
\bibitem [{Note2()}]{Note2}%
  \BibitemOpen
  \bibinfo {note} {The chiral chemical potential $\mu _5$ itself does not
  necessarily depend on UV properties of band structure. However, for the
  practical realization of nonzero $\mu _5$, we need some chirality source such
  as the chiral anomaly via external parallel electromagnetic fields $\protect
  \bm {E}\cdot \protect \bm {B}$. Then, the competition between the chirality
  pumping and chirality relaxation results in nonvanishing $\mu _5\propto \tau
  _5\protect \bm {E}\cdot \protect \bm {B}$ with the chirality relaxation time
  $\tau _5$. Therefore, $\mu _5$ indeed depends on UV properties because $\tau
  _5$ is determined by chirality relaxation mechanisms such as chirality
  flipping scattering and impurity scattering. It is very hard to theoretically
  predict $\tau _5$, which highly depends on microscopic nature of individual
  materials.}\BibitemShut {Stop}%
\bibitem [{\citenamefont {Chen}\ \emph {et~al.}(2014)\citenamefont {Chen},
  \citenamefont {Son}, \citenamefont {Stephanov}, \citenamefont {Yee},\ and\
  \citenamefont {Yin}}]{chen2014lorentz}%
  \BibitemOpen
  \bibfield  {author} {\bibinfo {author} {\bibfnamefont {J.-Y.}\ \bibnamefont
  {Chen}}, \bibinfo {author} {\bibfnamefont {D.~T.}\ \bibnamefont {Son}},
  \bibinfo {author} {\bibfnamefont {M.~A.}\ \bibnamefont {Stephanov}}, \bibinfo
  {author} {\bibfnamefont {H.-U.}\ \bibnamefont {Yee}}, \ and\ \bibinfo
  {author} {\bibfnamefont {Y.}~\bibnamefont {Yin}},\ }\href
  {http://link.aps.org/doi/10.1103/PhysRevLett.113.182302} {\bibfield
  {journal} {\bibinfo  {journal} {Phys. Rev. Lett.}\ }\textbf {\bibinfo
  {volume} {113}},\ \bibinfo {pages} {182302} (\bibinfo {year}
  {2014})}\BibitemShut {NoStop}%
\bibitem [{\citenamefont {Ishizuka}\ \emph {et~al.}(2016)\citenamefont
  {Ishizuka}, \citenamefont {Hayata}, \citenamefont {Ueda},\ and\ \citenamefont
  {Nagaosa}}]{Ishizuka:2016dvm}%
  \BibitemOpen
  \bibfield  {author} {\bibinfo {author} {\bibfnamefont {H.}~\bibnamefont
  {Ishizuka}}, \bibinfo {author} {\bibfnamefont {T.}~\bibnamefont {Hayata}},
  \bibinfo {author} {\bibfnamefont {M.}~\bibnamefont {Ueda}}, \ and\ \bibinfo
  {author} {\bibfnamefont {N.}~\bibnamefont {Nagaosa}},\ }\href {\doibase
  10.1103/PhysRevLett.117.216601} {\bibfield  {journal} {\bibinfo  {journal}
  {Phys. Rev. Lett.}\ }\textbf {\bibinfo {volume} {117}},\ \bibinfo {pages}
  {216601} (\bibinfo {year} {2016})}\BibitemShut {NoStop}%
\bibitem [{\citenamefont {Kharzeev}\ \emph {et~al.}()\citenamefont {Kharzeev},
  \citenamefont {Kikuchi},\ and\ \citenamefont {Meyer}}]{Kharzeev:2016mvi}%
  \BibitemOpen
  \bibfield  {author} {\bibinfo {author} {\bibfnamefont {D.}~\bibnamefont
  {Kharzeev}}, \bibinfo {author} {\bibfnamefont {Y.}~\bibnamefont {Kikuchi}}, \
  and\ \bibinfo {author} {\bibfnamefont {R.}~\bibnamefont {Meyer}},\
  }\href@noop {} {\ }\Eprint {http://arxiv.org/abs/1610.08986}
  {arXiv:1610.08986 [cond-mat.mes-hall]} \BibitemShut {NoStop}%
\bibitem [{\citenamefont {Huang}\ \emph {et~al.}(2016)\citenamefont {Huang},
  \citenamefont {Xu}, \citenamefont {Belopolski}, \citenamefont {Lee},
  \citenamefont {Chang}, \citenamefont {Chang}, \citenamefont {Wang},
  \citenamefont {Alidoust}, \citenamefont {Bian}, \citenamefont {Neupane},
  \citenamefont {Sanchez}, \citenamefont {Zheng}, \citenamefont {Jeng},
  \citenamefont {Bansil}, \citenamefont {Neupert}, \citenamefont {Lin},\ and\
  \citenamefont {Hasan}}]{huang2016new}%
  \BibitemOpen
  \bibfield  {author} {\bibinfo {author} {\bibfnamefont {S.-M.}\ \bibnamefont
  {Huang}}, \bibinfo {author} {\bibfnamefont {S.-Y.}\ \bibnamefont {Xu}},
  \bibinfo {author} {\bibfnamefont {I.}~\bibnamefont {Belopolski}}, \bibinfo
  {author} {\bibfnamefont {C.-C.}\ \bibnamefont {Lee}}, \bibinfo {author}
  {\bibfnamefont {G.}~\bibnamefont {Chang}}, \bibinfo {author} {\bibfnamefont
  {T.-R.}\ \bibnamefont {Chang}}, \bibinfo {author} {\bibfnamefont
  {B.}~\bibnamefont {Wang}}, \bibinfo {author} {\bibfnamefont {N.}~\bibnamefont
  {Alidoust}}, \bibinfo {author} {\bibfnamefont {G.}~\bibnamefont {Bian}},
  \bibinfo {author} {\bibfnamefont {M.}~\bibnamefont {Neupane}}, \bibinfo
  {author} {\bibfnamefont {D.}~\bibnamefont {Sanchez}}, \bibinfo {author}
  {\bibfnamefont {H.}~\bibnamefont {Zheng}}, \bibinfo {author} {\bibfnamefont
  {H.-T.}\ \bibnamefont {Jeng}}, \bibinfo {author} {\bibfnamefont
  {A.}~\bibnamefont {Bansil}}, \bibinfo {author} {\bibfnamefont
  {T.}~\bibnamefont {Neupert}}, \bibinfo {author} {\bibfnamefont
  {H.}~\bibnamefont {Lin}}, \ and\ \bibinfo {author} {\bibfnamefont {M.~Z.}\
  \bibnamefont {Hasan}},\ }\href {\doibase 10.1073/pnas.1514581113} {\bibfield
  {journal} {\bibinfo  {journal} {Proc. Natl. Acad. Sci. USA}\ }\textbf
  {\bibinfo {volume} {113}},\ \bibinfo {pages} {1180} (\bibinfo {year}
  {2016})}\BibitemShut {NoStop}%
\bibitem [{\citenamefont {Chang}\ and\ \citenamefont
  {Yang}(2015{\natexlab{b}})}]{PhysRevB.92.205201}%
  \BibitemOpen
  \bibfield  {author} {\bibinfo {author} {\bibfnamefont {M.-C.}\ \bibnamefont
  {Chang}}\ and\ \bibinfo {author} {\bibfnamefont {M.-F.}\ \bibnamefont
  {Yang}},\ }\href {\doibase 10.1103/PhysRevB.92.205201} {\bibfield  {journal}
  {\bibinfo  {journal} {Phys. Rev. B}\ }\textbf {\bibinfo {volume} {92}},\
  \bibinfo {pages} {205201} (\bibinfo {year} {2015}{\natexlab{b}})}\BibitemShut
  {NoStop}%
\bibitem [{\citenamefont {Eguchi}\ \emph {et~al.}(1980)\citenamefont {Eguchi},
  \citenamefont {Gilkey},\ and\ \citenamefont
  {Hanson}}]{eguchi1980gravitation}%
  \BibitemOpen
  \bibfield  {author} {\bibinfo {author} {\bibfnamefont {T.}~\bibnamefont
  {Eguchi}}, \bibinfo {author} {\bibfnamefont {P.~B.}\ \bibnamefont {Gilkey}},
  \ and\ \bibinfo {author} {\bibfnamefont {A.~J.}\ \bibnamefont {Hanson}},\
  }\href {http://www.sciencedirect.com/science/article/pii/0370157380901301}
  {\bibfield  {journal} {\bibinfo  {journal} {Phys. Rep.}\ }\textbf {\bibinfo
  {volume} {66}},\ \bibinfo {pages} {213 } (\bibinfo {year}
  {1980})}\BibitemShut {NoStop}%
\bibitem [{\citenamefont {Nakahara}(2003)}]{nakahara2003geometry}%
  \BibitemOpen
  \bibfield  {author} {\bibinfo {author} {\bibfnamefont {M.}~\bibnamefont
  {Nakahara}},\ }\href
  {https://www.crcpress.com/Geometry-Topology-and-Physics-Second-Edition/Nakahara/p/book/9780750306065}
  {\emph {\bibinfo {title} {Geometry, Topology and Physics}}},\ Graduate
  student series in physics\ (\bibinfo  {publisher} {CRC Press},\ \bibinfo
  {year} {2003})\BibitemShut {NoStop}%
\end{thebibliography}%

\end{document}